\documentclass[journal]{IEEEtran}
\pdfoutput=1
\ifCLASSINFOpdf
\else
\fi

\usepackage{url}            
\usepackage{booktabs}       
\usepackage{amsfonts}       
\usepackage{nicefrac}       
\usepackage{microtype}      
\usepackage[space]{grffile} 
\usepackage{subcaption}
\usepackage{graphicx}
\usepackage{amsmath}
\usepackage{cite}
\usepackage{multirow}
\usepackage{enumitem}
\usepackage{xcolor}
\usepackage{amssymb}

\begin{document}
%
\title{Weakly-supervised Visual Instrument-playing Action Detection in Videos}

%
%
%


\author{
  Jen-Yu~Liu, 
  Yi-Hsuan~Yang,
  Shyh-Kang~Jeng
\thanks{J.-Y. Liu (contact: ciaua@citi.sinica.edu.tw) and S.-K. Jeng are with Graduate Institute of Electrical Engineering, National Taiwan University, Taiwan}%
\thanks{J.-Y. Liu and Y.-H. Yang are with CITI, Academia Sinica, Taiwan}}

\maketitle


\begin{abstract}
Instrument playing is among the most common scenes in music-related videos, which represent nowadays one of the largest sources of online videos. In order to understand the instrument-playing scenes in the videos, it is important to know what instruments are played, when they are played, and where the playing actions occur in the scene. While audio-based recognition of instruments has been widely studied, the visual aspect of the music instrument playing remains largely unaddressed in the literature. One of the main obstacles is the difficulty in collecting annotated data of the action locations for training-based methods. To address this issue, we propose a weakly-supervised framework to find when and where the instruments are played in the videos. We propose to use two auxiliary models, a sound model and an object model, to provide supervisions for training the instrument-playing action model. The sound model provides temporal supervisions, while the object model provides spatial supervisions. They together can simultaneously provide temporal and spatial supervisions. The resulted model only needs to analyze the visual part of a music video to deduce which, when and where instruments are played. We found that the proposed method significantly improves the localization accuracy. We evaluate the result of the proposed method temporally and spatially on a small dataset (totally 5,400 frames) that we manually annotated.

\end{abstract}


%
\IEEEpeerreviewmaketitle

\ifCLASSOPTIONcaptionsoff
  \newpage
\fi

%
%

\section{Introduction} \label{sec:intro}

With the popularity of social media and online sharing, people are sharing a large amount of videos online every day. These videos often contain human activities, so human actions or movements are informative components in these videos. Therefore, automatically recognizing the types of the actions and locating the actions in videos can help understand and retrieve videos \cite{zhou2015}. This task is often called ``action detection'' \cite{zhou2015, shi2017, bojanowski2014, simonyan2014, ng2015, jiang2015, feichtenhofer2016, huang2016}. For fully-supervised learning approaches, detailed temporal and spatial annotations of actions are usually needed for action detection. However, these annotations are difficult to acquire because the labeling is labor-intensive and time-consuming \cite{zhou2015}. In recent years, researchers have proposed various strategies to alleviate this issue \cite{zhou2015, bojanowski2014, huang2016}.

We observe that the objects and sounds in several types of videos might also be used to alleviate this issue. In a large amount of videos, both the objects and sounds signify the key points of the actions. Examples include videos with instrument playing \cite{li2017}, videos with violent content \cite{hu2016}, and sport videos \cite{cricri2014}. For example, when we hear a guitar solo and see a musician holding a guitar in a video, it is pretty likely that the guitar solo comes from the musician's playing actions. The hitting actions in ball games often contain the acting objects, such as feet, hands, rackets, or bats, as well as the accompanied sounds of hitting. This relationship between actions, objects, and sounds provides an opportunity to infer the appearance of actions from objects and sounds. 

Specifically, from the actions in the videos where sounds and objects signify the key points of actions, we observe the following two common properties:
\begin{description}
\item [Action-in-object] The spatial location of an action is close to (e.g. at the border or within) the spatial location of objects (e.g., instruments, bats, balls, or weapons).
\item [Action-making-sound] A specific type of actions is associated with a specific type of sounds that the actions make.
\end{description}
Action-in-object together with the region of the objects in the scene may give us clues regarding where the actions occur spatially, while action-making-sound together with the temporal activation of the sounds in the video frames may help us temporally locate the actions. In contrast to annotated action data, annotated data of objects and annotated data of sounds are easier to acquire. Therefore, we propose to train a sound model specifying when the actions occur and train an object model telling us where the objects are. These two auxiliary models act as teachers to inform the action model when and where to pay attention to. We feed only the motion information (dense optical flows in this work) to the action model, so it is forced to learn when and where the actions occur by only motions with the help from the two auxiliary models. An interesting feature of this proposed framework is that it does not need annotated data of actions at all in the training process. We consider our proposed framework as a weakly-supervised learning one, because the model is trained to predict \emph{when} and \emph{where} the playing actions are in videos by using only information regarding \emph{whether} an instrument appears in a video clip in the training phase. The proposed framework is depicted in Fig. \ref{fig:sot} (Figs. \ref{fig:vt}, \ref{fig:ot}, and \ref{fig:st} are variants of the proposed framework that will be discussed in Section \ref{sec:method}).

We focus on the instrument-playing actions in music-related videos in this work. Music is one of the most popular types among online videos (ranked number two according to the study of Cheng \emph{et al.} \cite{cheng2013}), and instrument playing is among the most common scenes in these videos. For the audio aspect of instrument playing, automatic detection of instrument sounds has been widely studied in music information retrieval (MIR) \cite{essid2006, fu2011, giannoulis2014, han2017, rizzi2017}. It helps people understand the content of the music. However, the visual aspect of instrument playing remains largely unaddressed in literature. In addition to the sounds of instruments, the visual appearances of instruments and instrument-playing actions also provide us important information about the music-related videos. In order to understand music-related videos, we need to know which instruments are played, when the instruments are played, and where the playing actions occur in the scene.\footnote{And even \emph{how} the instruments are played---the gesture, the playing technique, the expression etc \cite{reboursiere2012}. We leave this as a topic of future research.} For example, in a video of a piano concert, the pianist may first walk into the scene, sit down, and then start to play the piano. In this case, the piano is not played until the pianist sits down and is ready. We may want to know when the playing begins, the relative position of the piano to the scene, the relative positions of the hands to the piano, etc. There are also attempts to model the audio and visual information jointly for music information retrieval tasks \cite{schindler2016, slizovskaia2017}. For example, Schindler \emph{et al.} investigated music genre classification by aggregating audio features and visual features together as the input features to a classifier \cite{schindler2016}. This approach could improve the input feature of the model, but cannot circumvent the lack of annotated data.

\begin{figure}[!t]
\centering
\includegraphics[width=0.95\columnwidth]{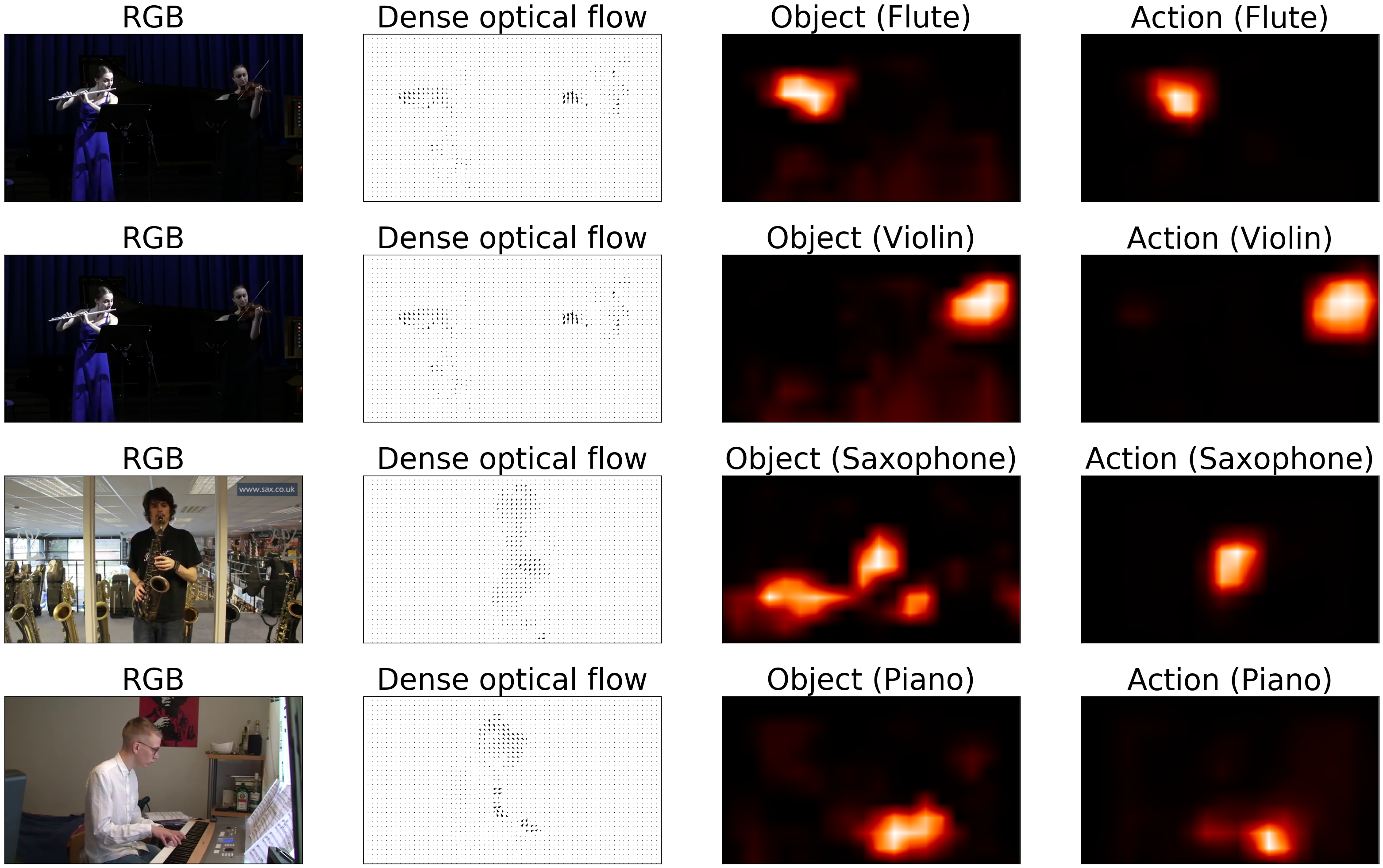}
\caption[Examples]{Examples of action detection result. The first column are frames from online videos, the second column are the dense optical flows of those frames, and the third and fourth columns are the results of object detection and action detection of our models, respectively.\footnotemark}
\label{fig:examples}
\end{figure}

\footnotetext{The left-most RGB snapshots are cropped from YouTube videos \url{https://www.youtube.com/watch?v=<d3J_aYbTaEE, _mpZqTZxJrU, BAieBB1yhfw>} uploaded by FluteMasters, Sax.co.uk, and Jaakko Kiuru, respectively. All of them are under Creative Commons license.}

In light of these observations, the goal of this work is to train a model to automatically pinpoint the instrument-playing actions temporally and spatially in videos with instrument-playing scenes without detailed annotations. In contrast to the abundance of annotated data available for either object recognition (including instruments), such as ImageNet\footnote{http://www.image-net.org/} \cite{ILSVRC15}, or sound recognition, such as AudioSet\footnote{https://research.google.com/audioset/}, we have no available dataset specifying the location of the playing actions in the scenes. Therefore, we can train the action model by utilizing the two properties mentioned above together with a trained sound model and a trained object model. We use the spatial locations of instrument objects and the temporal locations of the instrument sounds to help the detection of playing actions, but do not join the input features. In this way, we have a more flexible model that can work even if the audio is degraded due to factors such as environmental noises, audio track loss, or audio compression artifacts \cite{jang2016}. We human beings can guess if an instrument is played simply by the action, gesture, and the relative positions of hands or bows to instruments. Some examples of action detection result are shown in Fig. \ref{fig:examples}. We can see that there could be multiple types of instruments in the scene at the same time, and instrument and the playing actions do not always coincide.

Our contribution is three-fold. First, we propose a training framework to learn the temporal and spatial locations of the actions without detailed annotations by utilizing the object and the sound information (Section \ref{sec:method}). 
Furthermore, we find that we can utilize the object and sound information to further improve the result after the action model is trained by a simple yet effective method of model fusion (Section \ref{sec:fusion_method}). 
Second, although the proposed method does not require detailed location information in training, for the purpose of evaluation, we manually annotated totally 5,400 frames from 135 videos with detailed locations of instrument-playing actions (Section \ref{sec:dataset}). Third, we conduct comprehensive experiments to investigate the effects of different components in the framework (Section \ref{sec:exp}). We also analyze the action patterns the neural network learns for each instrument. Furthermore, the code and the manually-annotated data are made publicly available (https://github.com/ciaua/InstrumentPlayingDetection) for further investigation and for reproducibility.


\section{Related work} \label{sec:related}
Our work stands as an intersection of weakly-supervised object detection, action detection in videos, and instrument sound detection. We survey the most related works in these fields and put our work in the context.

\subsection{Image and video classification}

Zhou et al. identified the difficulty in acquiring action annotations for action detection and proposed a way to estimate the temporal and spatial extents of the actions \cite{zhou2015}. They proposed a trajectory split-and-merge algorithm to first segment the background and the foreground moving objects by using dense optical flows, and then they used the segmentation information to derive the temporal and spatial extents of the actions. Then, they used a latent SVM to classify these segmented patches and locate the actions. We share a similar goal to derive the temporal and spatial extents in our proposed framework, but we investigate utilizing two other modalities to estimate the extents, instead of using the dense optical flows.

Oquab \emph{et al.} proposed to use fully-convolutional neural networks (FCNs) to realize weakly-supervised learning for images \cite{oquab2015}. By replacing the fully-connected layers in conventional convolutional neural networks (CNNs) \cite{lecun1998, krizhevsky2012} with fully-convolutional layers, the model produces an output map that indicates the activation values at different locations. We use this method to do spatial weakly-supervised learning for both action detection and object detection in this work.

There have been several studies on weakly-supervised object detection or segmentation. Hartmann \emph{et al.} used support vector machine (SVM) \cite{cortes1995} for weakly-supervised object segmentation in videos \cite{hartmann2012}. Liu \emph{et al.} used a nearest neighbor-based method to perform weakly-supervised object segmentation in videos \cite{liu2014}. Prest \emph{et al.} used motion cues to produce candidates of temporal tubes that locate a moving object and trained the object detector with a subset of the tubes \cite{prest2012}.

Bojanowski \cite{bojanowski2014} and Huang \emph{et al.} \cite{huang2016} tackled the problem of weakly-supervised action detection. In their study, they only know the sequence of actions and they have to align the actions with the frames in a video clip. They proposed different ways to align the action sequence. Our work is different from theirs in two ways. First, we use auxiliary sound and object models to learn to assign labels to video frames, instead of based on the sequence of labels assigned by human. Second, they only attempt to predict the labels temporally but not spatially. 

Simonyan \emph{et al.} proposed a two-stream framework for action detection by using an object stream and an action stream \cite{simonyan2014}. They experimented with fusing the two streams either by averaging the output scores of the two models or by using SVM to do the final classification. Feichtenhofer \emph{et al.} extended Simonyan's work by using different ways of model fusion \cite{feichtenhofer2016}, and Ng \emph{et al.} extended Simonyan's work by incorporating information across longer period of time through temporal pooling and LSTM \cite{ng2015}. Our method also contains multiple streams. However, we use FCNs for all the three streams instead of the conventional CNNs because we want not only to classify the videos but also to locate the instruments, the actions, and the sounds. Furthermore, the models are fused only after they are separately trained.


Our method is also related to supervision transfer introduced by Gupta \emph{et al.} \cite{gupta2016}. Given two learning tasks where the task 1 has large annotated data while the task 2 does not, they proposed to use the output of a middle layer in the well-trained network in task 1 to provide supervision to a middle layer of the network in task 2. In this way, the supervision is transferred. In this work, we also want to seek for more supervisions to the instrument-playing actions from two other modalities, but we provide the supervisions directly in the output layers by the physical relationships of the three modalities that are indicated by the two observations stated in Section \ref{sec:intro}. The temporal and spatial supervisions are also exploited in addition to the instance-level label supervision in this work.

In recent years, unsupervised learning, weakly-supervised, and semi-supervised learning have received lots of attentions in video processing \cite{aytar2016, canziani2017, arandjelovic2017, aytar2017}. This trend is partly due to the lack of supervisory signals in videos, but it is also because the multi-modal nature of videos and the temporal continuity of videos provide a good environment for learning feature representations by the dependencies between modalities or between frames without external supervisions. For example, Aytar \emph{et al.} \cite{aytar2016} and Arandjelovi{\'{c}} \emph{et al.} \cite{arandjelovic2017} proposed to match the audio and visual information to unsupervisedly learn features from a large amount of videos and use only a few labeled data for training a classifier based on the learned features. Aytar \emph{et al.} \cite{aytar2017} further included text in addition to the audio and visual information for feature learning. In contrast to the aforementioned multi-modal approaches, Canziani \emph{et al.} \cite{canziani2017} proposed a CortexNet framework to learn features by matching neighboring frames in videos. Similar to these works, our proposed framework also represents an attempt to increase supervisory signals by utilizing multiple modalities of videos for a challenging action detection task.

\subsection{Audio classification}

Instrument recognition has been an active research topic in MIR. Essid \emph{et al.} extracted various audio features and applied hierarchical clustering and SVM for instrument recognition \cite{essid2006}. Han \emph{et al.} proposed a CNN structure to recognize the predominant instrument in music \cite{han2017}. Slizovskaia \emph{et al.} used both audio and visual features as the input and applied CNNs for the task of instrument recognition \cite{slizovskaia2017}. The goal of these works is to recognize the instrument sounds with audio or audio-visual information as the input. In contrast, our goal is to detect the instrument-playing actions at frame level by the visual cues in a video.



In recent years, the multimedia and MIR community starts to address the difficulty in collecting training data for frame-level predictions. Kumar \emph{et al.} investigated the problem of audio event detection with SVM and neural networks with weak-labeled data \cite{kumar2016}. Schl{\"{u}}ter utilized saliency maps to iteratively train a model that can recognize singing voices in the frame level \cite{schluter2016learning}. Liu \emph{et al.} applied FCNs to a general music auto-tagging problem so that the model detects various music-related properties in the frame level, including genres, instruments, vocals, etc \cite{liu2016}. In this work, we also utilize an FCN model to derive the frame-level instrument sound predictions. However, we use the frame-level instrument sound predictions as the training target for the visual action model, instead of as the end product itself.


\begin{figure*}
\centering
\begin{subfigure}{0.4\textwidth}
 \centering
 \includegraphics[height=5.5cm]{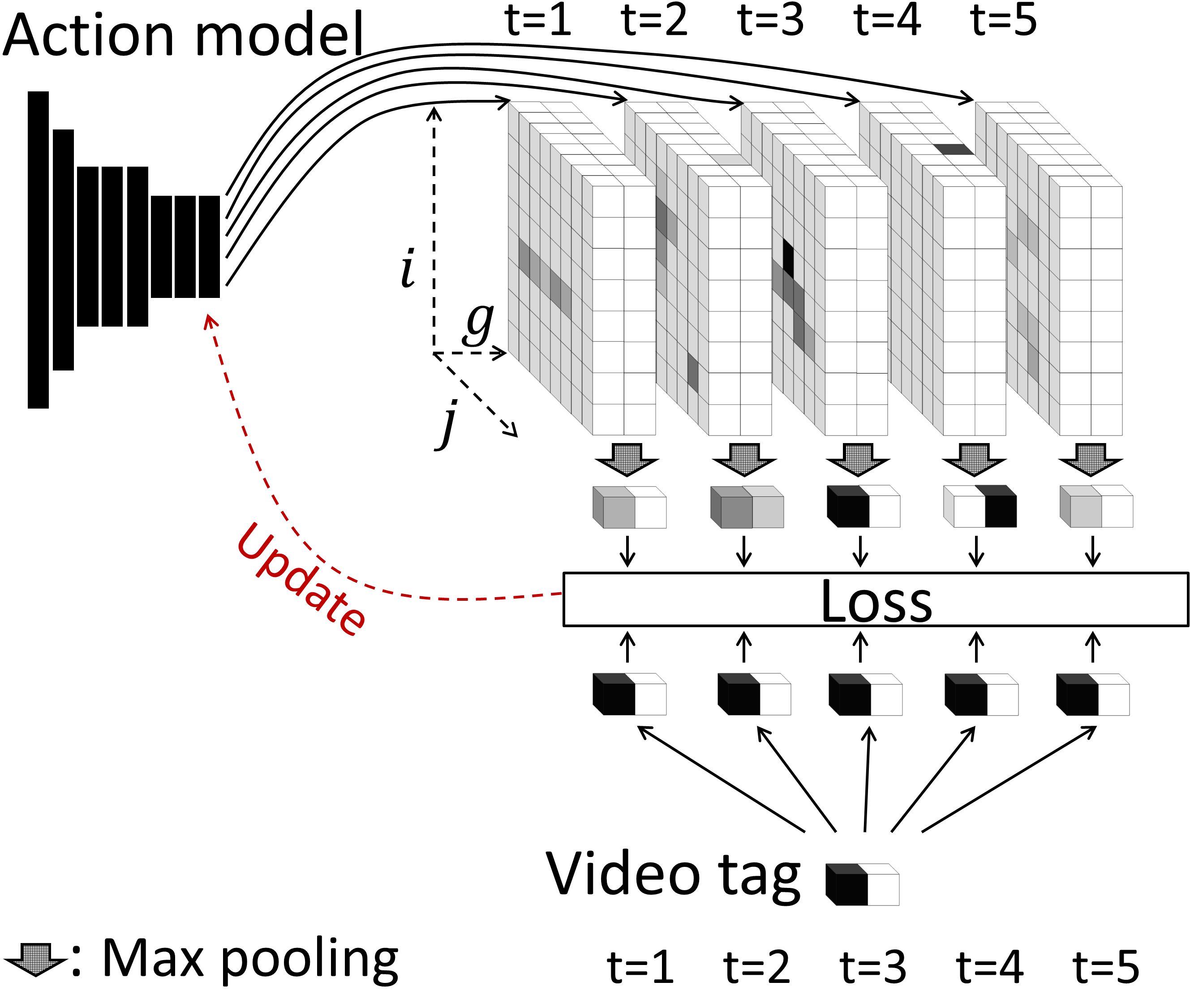}
  \caption{Video tag as target (VT): $loss(\max\limits_{i, j} A_{g, t, i, j}, V_{g})$}
  \label{fig:vt}
\end{subfigure}%
\begin{subfigure}{0.6\textwidth}
 \centering
 \includegraphics[height=5.5cm]{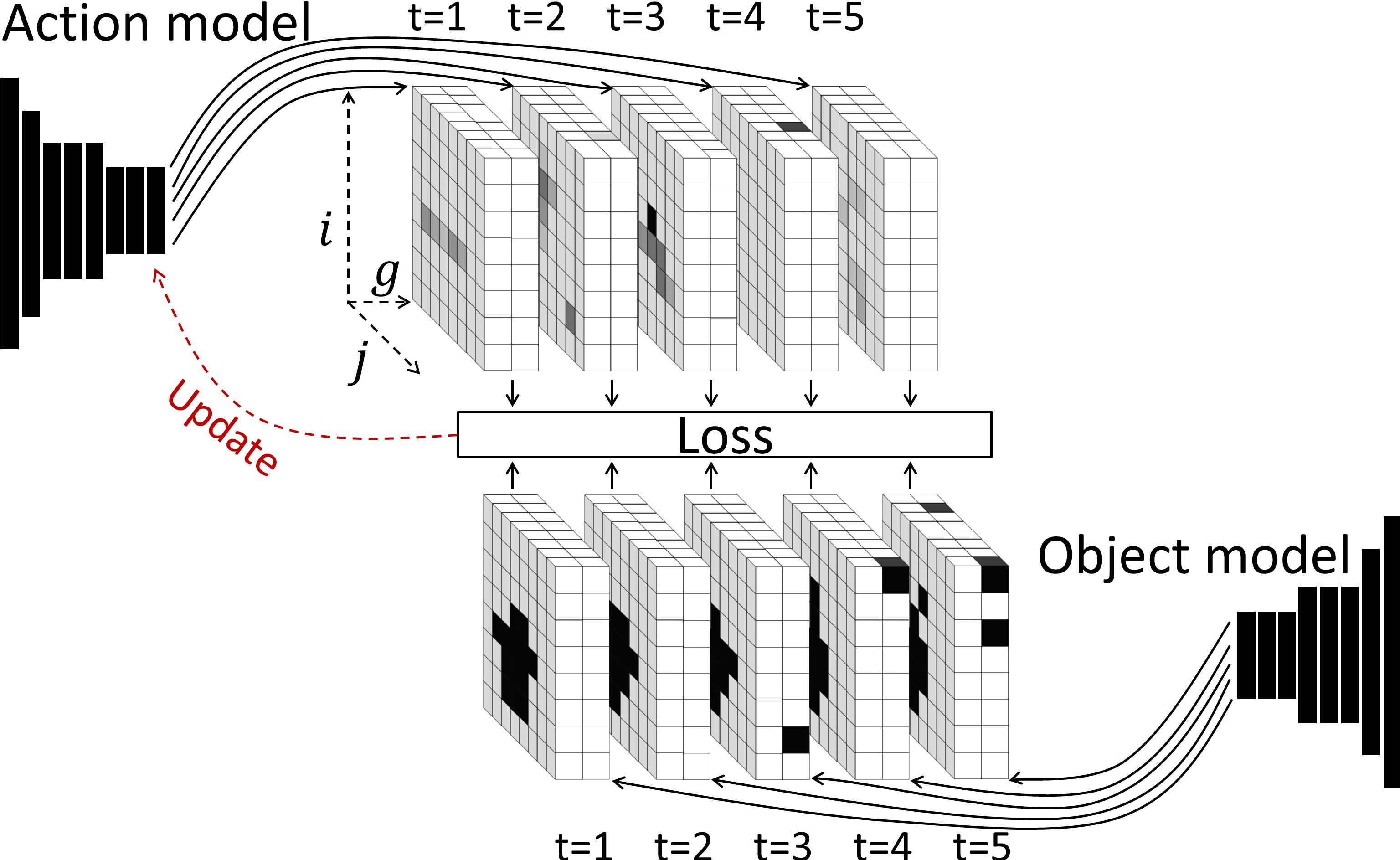}
  \caption{Object as target (OT): $loss(A_{g, t, i, j}, \hat{O}^u_{g, t, i, j})$}
  \label{fig:ot}
\end{subfigure}
\par\bigskip
\begin{subfigure}{0.37\textwidth}
 \centering
 \includegraphics[height=5.5cm]{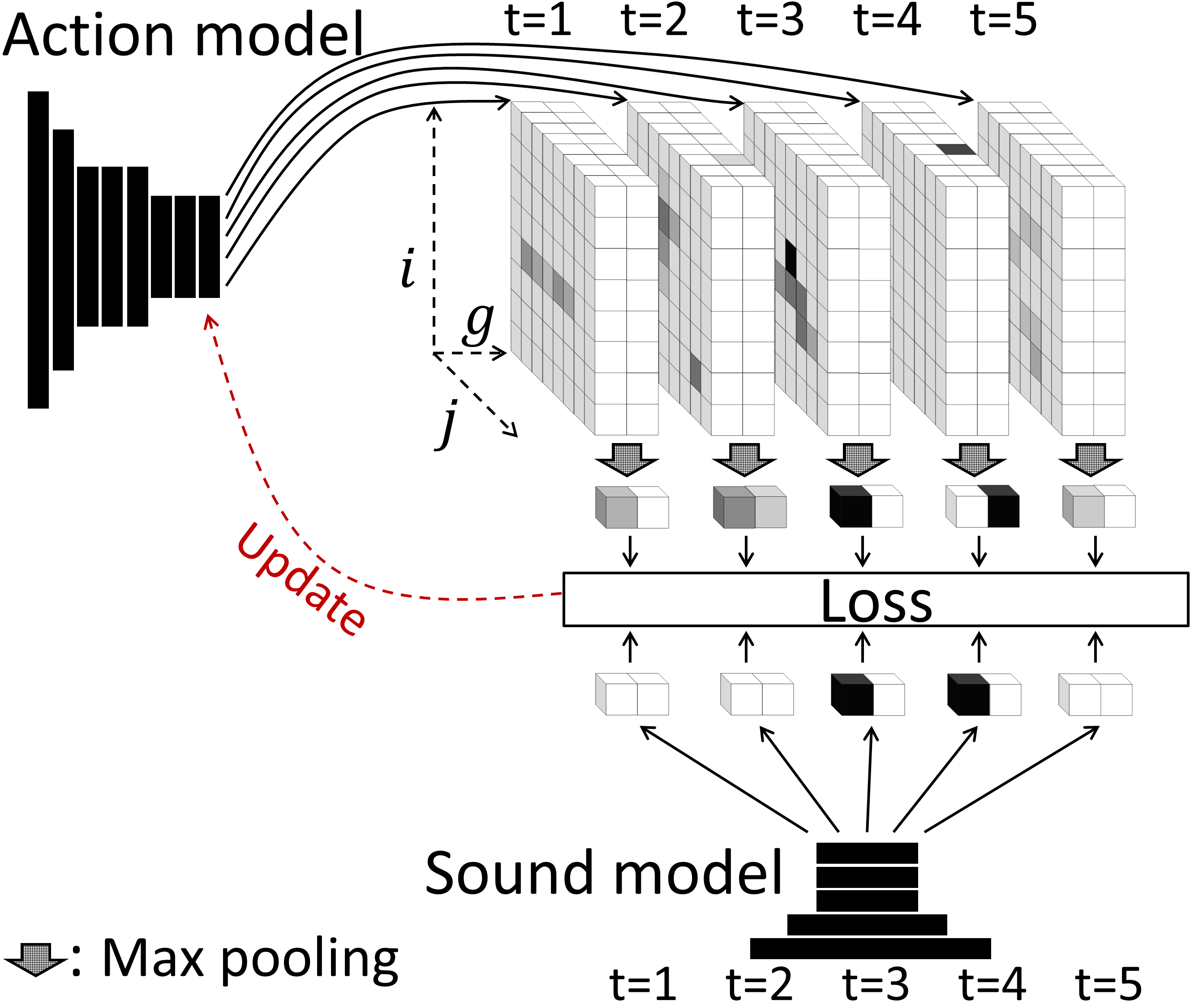}
  \caption{Sound as target (ST): $loss(\max\limits_{i, j} A_{g, t, i, j}, \hat{S}^v_{g, t})$}
  \label{fig:st}
\end{subfigure}%
\begin{subfigure}{0.63\textwidth}
 \centering
 \includegraphics[height=5.5cm]{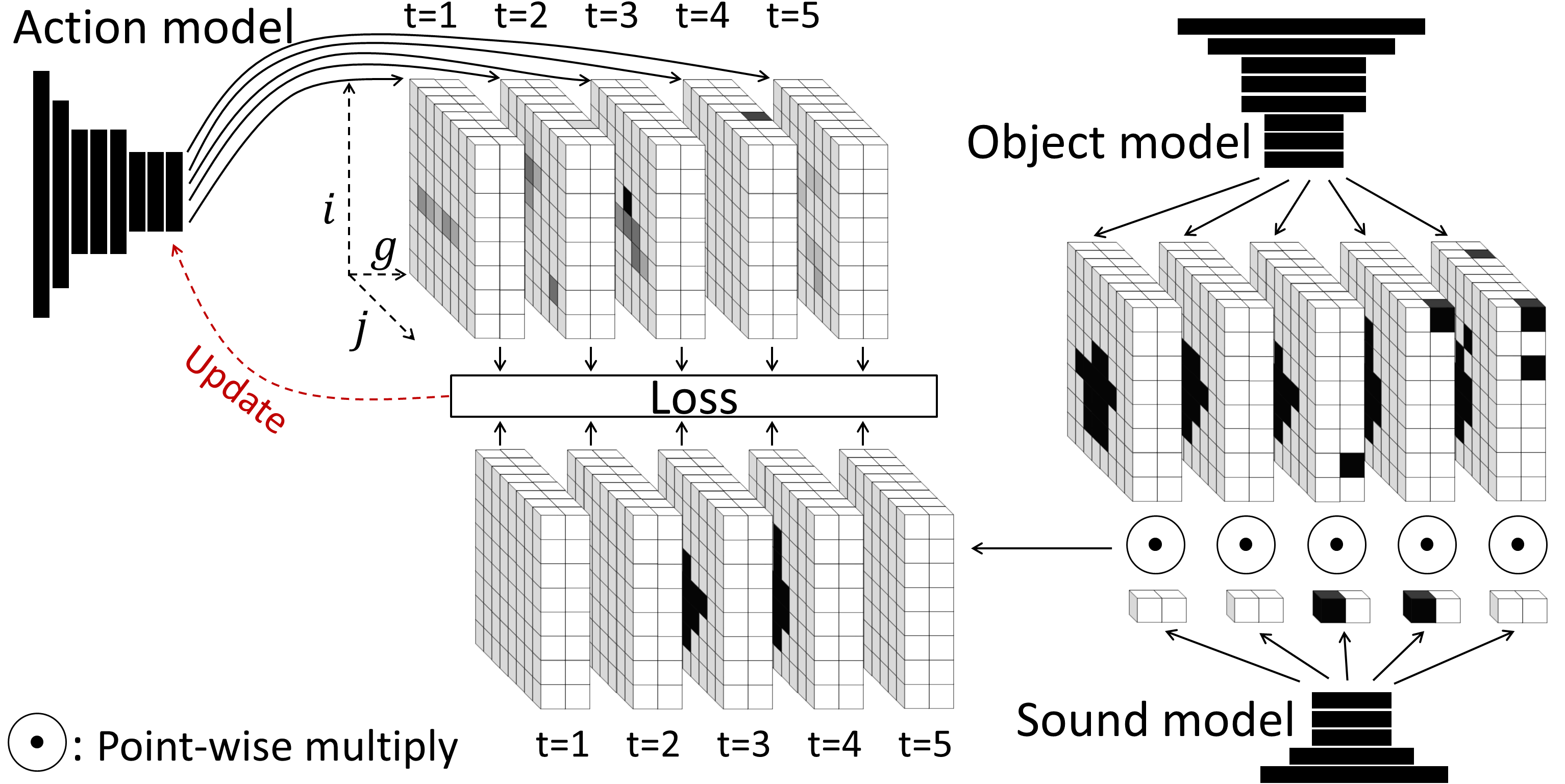}
  \caption{Sound$\times$Object as target (SOT): $loss(A_{g, t, i, j}, \hat{S}^v_{g, t}\hat{O}^u_{g, t, i, j})$}
  \label{fig:sot}
\end{subfigure}

\caption{Four levels of supervisions. $g$ is the instrument index, $t$ is the temporal index, and $(i, j)$ is the spatial coordinate in the output map. $[V_g]$ is the vector of the video tags, $[\hat{O}^u_{g, t, i, j}]$ is the binarized output of the object model, $[\hat{S}^v_{g, t}]$ is the binarized output of the sound model, and $[A_{g, t, i, j}]$ is the output of the action model. Darker shade represents higher activation where activation values are between 0 and 1. Each cuboid in the figure represents an output tensor produced by a model at a given time $t$, and the three axes of a cuboid represent the instrument index $g$ and the spatial coordinate $(i, j)$, respectively. The $t$ above or below a cuboid represents its temporal index.}
\label{fig:framework}
\end{figure*}

\section{Proposed method}  

In what follows, we firstly describe the proposed weakly-supervised method by increasing supervisions in the process of training the action model by information from other modalities. Then, we describe how we fuse information from other modalities, after the action model has been trained.

\subsection{Increasing supervisions for training the action model}\label{sec:method}

There could be various movements in the body or the instrument when a musician is playing an instrument. In this work, ``playing actions'' refer to the movements or actions that are responsible for making the instrument sounds in the scene. For example, the playing actions of flute include the movements of hands and mouths, while the playing actions of guitar include the movements of the fretting hand and the picking hand. The goal is to determine whether playing actions are present, which instruments correspond to the playing actions, and where the playing actions are in the scene.

An intuitive way to build an instrument-playing action detector is to utilize the instrument tags accompanying the videos for model training. The tags can be either from a video dataset or from the titles of the videos on an online-streaming website. We may assume that each frame contains the playing actions of this tag $g$ somewhere in the scene. Therefore, while training the action model, we can apply a spatial max-pooling to the output of the action model and compare it with the tag of the video clip, $V=[V_g]$, by a loss function, that is, $loss(\max\limits_{i, j} A_{g, t, i, j}, V_{g})$, where $t$ is the temporal index, $(i, j)$ is the spatial coordinate,  and $A=[A_{g, t, i, j}]$ is the output of the action model. $V=[V_g]$ can be either a one-hot vector or with multiple positive entries, depending on the dataset. This weakly-supervised approach is similar to the approach of weakly-supervised object detection in images by Oquab \emph{et al.} \cite{oquab2015}. An illustration is shown in Fig. \ref{fig:vt}. An action model trained with this approach will be referred to as a VT model. 

In the context of action detection, the supervision provided by such a weakly-supervised approach may be poor, because the musician might not be playing the instrument all the time throughout the video. Furthermore, the action model needs to search through the entire scene for playing actions while there are potentially many movements in the scene that are irrelevant to the playing actions. 

A plausible way to improve the performance is to include information from different modalities. Similar to how Simonyan \emph{et al.} fused information from a still-image stream and an action stream in their two-stream action detection model \cite{simonyan2014}, we can fuse information from a still-image stream, an action stream, and an audio (sound) stream so that we have more information available. This could improve the performance, as we will show in Section \ref{sec:fusion}, but it does not deal with the problem of lacking supervisions itself.

Can we limit the search space for the action model and improve the supervision we have? We can achieve it by using the observation `action-in-object' introduced in Section \ref{sec:intro}, stating that the actions responsible to the instrument sound should occur in the region of the instrument. Assume we have a well-trained object model that can inform us the locations of instruments in the form of an output tensor $O=[O_{g, t, i, j}]$ whose value at a given location, $(i, j)$, specifying the confidence level of observing an instrument $g$ there. By binarizing $O$ with respect to a threshold $u$, that is,  
\begin{equation} 
\hat{O}^u_{g, t, i, j}=\begin{cases} 1 & \text{if $O_{g, t, i, j}\geq u$} \\ 0 & \text{if $O_{g, t, i, j}<u$} \end{cases},
\end{equation}
we can limit the search space to the region of the instrument by asking the action model to regard only the region of the instrument as positive, i.e., $loss(A_{g, t, i, j}, \hat{O}^u_{g, t, i, j})$. An illustration is shown in Fig. \ref{fig:ot}. A downside of this approach, however, is that the musician could just bring the instrument alongside without playing it. The appearance of the instrument does not necessarily imply the playing of the instrument at a given time. We refer to an action model trained with this approach as an OT model.

\begin{table*}[htbp]
\renewcommand{\arraystretch}{1.3}
\caption{Architecture of the sound model. It is an FCN adapted from the model proposed by the authors in their previous work \cite{liu2016}. There are three scales of input feature maps, and each of them has their own stack of early convolutions (Conv1 and Conv2). `RF' represents receptive field and `St' represents stride size.}
\label{tab:sound_structure}
\centering
\begin{tabular}{|c|c|c|c@{\hskip 4pt}c|c|c|c|c|}
\hline
	& \multicolumn{2}{c|}{Early convolutions} & &  & \multicolumn{3}{c|}{Late convolutions} & Global\\
\cline{2-3}
\cline{6-8}
Input ($\times$3 scales) & Conv1 ($\times$3 scales) & Conv2 ($\times$3 scales) & & & Conv3 & Conv4 & Conv5 ($S$) & pooling\\
\hline
\hline
128 channels	& Filter \#: 256  		& Filter \#: 256 &	\multirow{4}{*}{\rotatebox[origin=c]{90}{Concatenate}} & \multirow{4}{*}{\rotatebox[origin=c]{90}{3 scales}}	& Filter \#: 512		& Filter \#: 512		& Filter \#: 9			& Average \\
Log 	& RF: 5. Pad: 2. St: 1	& RF: 5. Pad: 2. St: 1 	& & & RF: 1. Pad: 0. St: 1	& RF: 1. Pad: 0. St: 1	& RF: 1. Pad: 0. St: 1	& pooling	 \\
mel-spectrogram		& Batch normalization	& Batch normalization & &	& Batch normalization	& Batch normalization	& 				& \\
			& Pool: 4			& Pool: 4 	&	&	& Dropout			& Dropout 			& Sigmoid function		& \\
\hline

\end{tabular}
\end{table*}

\begin{table*}[htbp]
\renewcommand{\arraystretch}{1.3}
\caption{Architecture of the action model and the object model. It is an FCN adapted from the CNN model VGG\_CNN\_M\_2048 used in \cite{chatfield2014}. `RF' represents receptive field and `St' represents stride size, and `LRN' represents local response normalization}
\label{tab:visual_structure}
\centering
\begin{tabular}{|c|c|c|c|c|c|c|c|c|c|}
\hline
	& \multicolumn{5}{c|}{Early convolutions} & \multicolumn{3}{c|}{Late convolutions} & Global \\
\cline{2-9}
Input & Conv1 & Conv2 & Conv3 & Conv4 & Conv5 & Conv6 & Conv7 & Conv8 ($O$, $A$) & pooling \\
\hline
\hline
Object		& Filter \#: 96  	& Filter \#: 256 	& Filter \#: 512	& Filter \#: 512	& Filter \#: 512	& Filter \#: 2048	& Filter \#: 1024	& Filter \#: 9 & Max \\
3 channels	& RF: 7x7		& RF: 5x5 		& RF: 3x3		& RF: 3x3		& RF: 3x3		& RF: 3x3		& RF: 1x1		& RF: 1x1 & pooling \\
\cline{1-1}
\cline{10-10}
Action		& Pad: 3. St: 2	& Pad: 2. St: 2	& Pad: 1. St: 1	&Pad: 1. St: 1	& Pad: 1. St: 1	& Pad: 0. St: 1	& Pad: 0. St: 1	& Pad: 0. St: 1 & None \\
10 channels	& LRN. Pool: 2	& Pool: 2 		& 			& 	 		& Pool: 2		& Dropout		& Dropout		& Sigmoid &  \\
\hline

\end{tabular}
\end{table*}


\begin{figure}[!t]
	\centering
	\includegraphics[width=0.95\columnwidth]{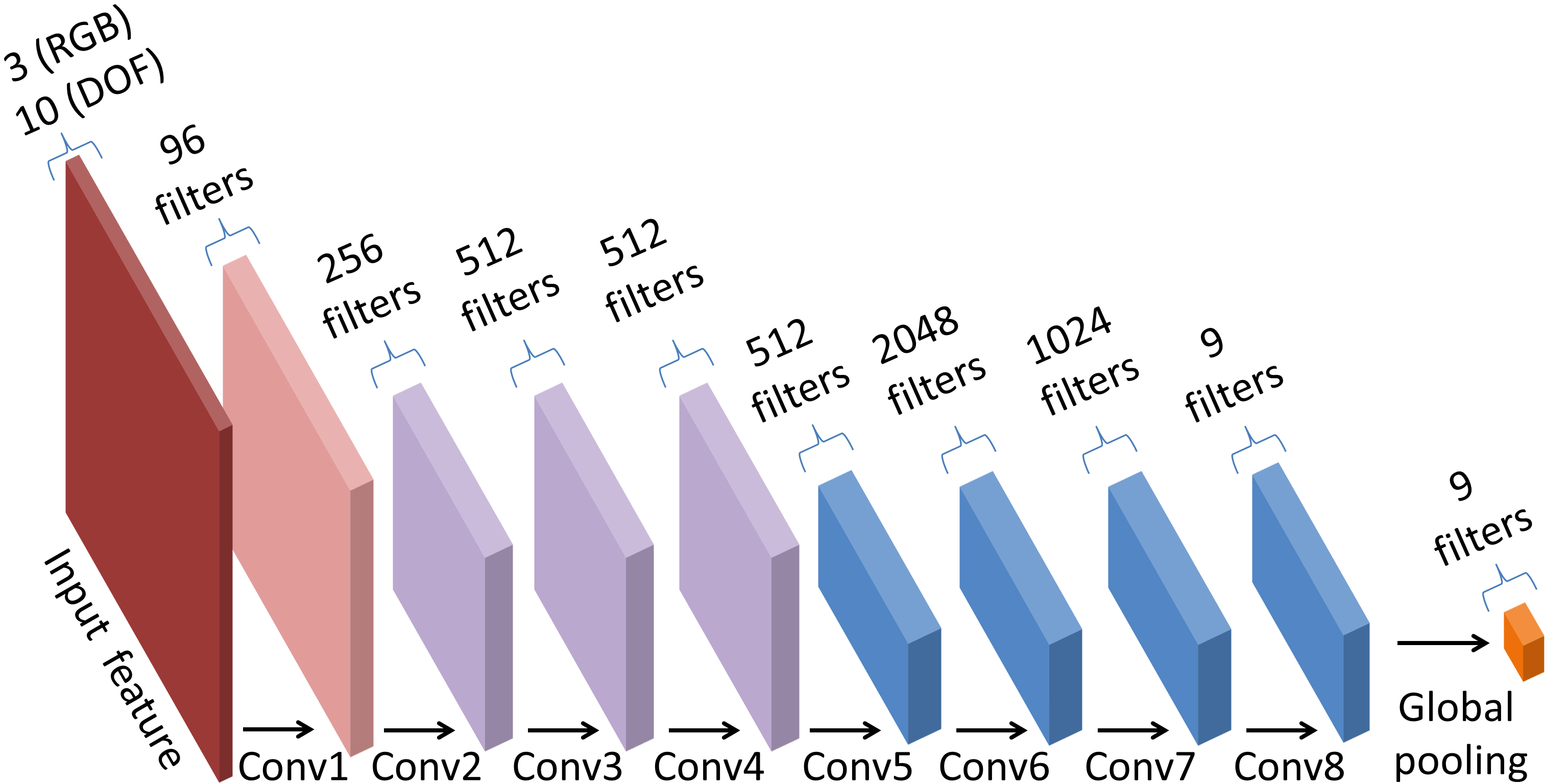}
	\caption{Architecture of the action and object models. Details of the model architecture are described in TABLE \ref{tab:visual_structure}.}
	\label{fig:architecture}
\end{figure}


Interestingly, we do have a way to acquire information about when an instrument is played. We can utilize the second observation `action-making-sound' introduced in Section \ref{sec:intro}, stating that an action of instrument playing would usually produce the sound of that instrument. Assume we have a well-trained sound model that can inform us if there are sounds of a specific instrument in a frame in the form of an output matrix $S=[S_{g, t}]$ specifying the confidence level of observing sounds of an instrument $g$ at a time frame $t$. By binarizing $S$ with respect to a threshold $v$, that is, 
\begin{equation}\label{eq:binary_sound}
\hat{S}^v_{g, t}=\begin{cases} 1 & \text{if $S_{g, t}\geq v$} \\ 0 & \text{if $S_{g, t}<v$} \end{cases},
\end{equation}
we can inform the action model to consider a frame as containing playing actions of an instrument only if this frame has sounds of the instrument. The corresponding loss for training is $loss(\max\limits_{i, j} A_{g, t, i, j}, \hat{S}^v_{g, t})$. An illustration is shown in Fig. \ref{fig:st}. An action model trained with this approach will be referred to as an ST model.

In order to have the benefits of both an OT model and an ST model, we can combine the output of the object model and the output of the sound model by point-wise multiplication, that is, $\hat{S}^v_{g, t}\hat{O}^u_{g, t, i, j}$. The corresponding training loss is $loss(A_{g, t, i, j}, \hat{S}^v_{g, t}\hat{O}^u_{g, t, i, j})$. An illustration is shown in Fig. \ref{fig:sot}. In this way, we inform an action model to search for playing actions only in the region containing the instrument in a frame containing the sounds of the instrument. We refer to an action model trained with this approach as an SOT model.

In summary, to circumvent the lack of annotated data, our proposed framework tries to acquire supervisions from other modalities. A trained object model provides spatial supervisions while a trained sound model provides temporal supervisions. Furthermore, we do not need the original clip-level tags anymore when we train an SOT action model. A list of the models used in this paper is presented in TABLE \ref{tab:modellist}.


\subsection{Fusion of different modality streams after model training} \label{sec:fusion_method}

In Section \ref{sec:method}, we have intentionally avoided including either still images or audio information into the input features of the action model so that the action model can be a standalone model without information from other modalities. However, we can still utilize the trained sound and object models to assist the trained action model when they are available. 

We adopt a straightforward fusion strategy by point-wisely multiplying the output layers of the models. Similar to the four levels of supervisions introduced in Section \ref{sec:method}, we can also have four levels of assistance from the sound and object models. 
\begin{description}
	\item [Action only] $[M_{g, t, i, j}]=[A_{g, t, i, j}]$
	\item [Action$\times$Object] $[M_{g, t, i, j}]=[A_{g, t, i, j}O_{g, t, i, j}]$
	\item [Action$\times$Sound] $[M_{g, t, i, j}]=[A_{g, t, i, j}S_{g, t}]$
	\item [Action$\times$Object$\times$Sound] $[M_{g, t, i, j}]=[A_{g, t, i, j}O_{g, t, i, j}S_{g, t}]$
\end{description}

When we use the last fusion formula, it is equivalent to saying that a playing action of a given instrument $g$ occurs at location $(i, j)$ at time $t$ if there are playing movements of the instrument $g$ at location $(i, j)$ at time $t$ AND there are the appearances of the instrument $g$ at location $(i, j)$ at time $t$ AND there are sounds of the instrument $g$ at time $t$.


\section{Experimental setup} \label{sec:setup}

\subsection{Models}

\subsubsection{Sound model: FCN trained with audios}
The sound model produces frame-level instrument sound predictions. We use Equation (\ref{eq:binary_sound}) to binarize the raw output, which is then used as part of the target to train the action model.


We implement a fully-convolutional network as the sound model that performs 1D convolutions. It is similar to the one proposed by Liu \emph{et al.} \cite{liu2016}, which has led to state-of-the-art result in frame-level music auto-tagging. There are mainly four differences between our model and the previous one \cite{liu2016}. First of all, we use different number of filters for the early convolution layers, which slightly improves the performance. Second, we use three instead of six feature maps for efficiency. Third, we do not use the Gaussian filters in the output because we want a sharper prediction. We find that the Gaussian filters could improve the frame-level predictions but will also blur the boundaries. Fourth, we apply a batch-normalization layer \cite{ioffe2015} after every convolution layer except for the output layer, to make the training process more stable with respect to parameter initialization. Each scale of the input feature maps is processed by its own early convolution layers, and then the outputs from three Conv2 layers are concatenated. The architecture is shown in TABLE \ref{tab:sound_structure}. We use the output of the Conv5 layer as the output, $S$, of a sound model.

\subsubsection{Object model}

The object model is implemented with an FCN to locate the instruments in the scene. Convolution layers process input feature maps locally so the location information is maintained. Therefore, an FCN can be used to locate objects in images, as proposed by Oquab \emph{et al.} \cite{oquab2015}. It takes still RGB images as the input. We use an architecture similar to VGG\_CNN\_M\_2048 \cite{chatfield2014} but with some modifications, shown in TABLE \ref{tab:visual_structure} and Fig. \ref{fig:architecture}. Importantly, all the fully-connected layers are replaced with convolution layers. We use the output of the Conv8 layer as the output, $O$, of an object model.

It has been shown that utilizing the pre-trained parameters of a good model can improve the performance for image recognition \cite{simonyan2014,long2015}. Following this light, the object model is modified from the VGG\_CNN\_M\_2048\footnote{We use parameters from \url{https://gist.github.com/ksimonyan/78047f3591446d1d7b91}} in \cite{chatfield2014}. Although there are other models that have better accuracy, we choose this one for it has reasonably good performance and for it can fit into the GPU of our computing machine.

One possible way to use the pre-trained CNN model is to directly transform the fully-connected layers in the CNN to convolution layers, as is done by Long \emph{et al.} \cite{long2015}. However, we have found that this direct conversion results in a poor localization model because, in the original setting of VGG\_CNN\_M\_2048, it takes an image of size 224x224 as the input and produces an output of size 6x6 right before the fully-connected layers, which is too coarse for the purpose of object localization. To cope with this issue, we use only the convolution layers from VGG\_CNN\_M\_2048 and discard all the fully-connected layers. We then append three convolution layers with a smaller $3\times 3$ receptive field as shown in TABLE \ref{tab:visual_structure}. We train the object model by regarding all the frames in a video clip of an instrument label as positive instances of the instrument, the same as the VT supervision for the action model in Fig. \ref{fig:vt}. 

A discussion about possible alternative models for the object model and some experimental results can be found in Appendix \ref{sec:fasterrcnn}.



\subsubsection{Action model}
The action model is implemented with an FCN to capture the actions of instrument-playing. It takes a stack of dense optical flows \cite{simonyan2014, farneback2003} as the input. We expect the action model to locate the playing actions in the scene. The architecture is the same as the object model but with a different number of input channels. 
It is trained from scratch without pre-training. 
We use the output of the Conv8 layer as the output, $A$, of an action model.

\subsection{Features}

A sampling rate of 16,000 and hop size of 512 are used to extract log mel-spectrograms from audios, and we use 3 scales of log mel-spectrograms with window size 512, 2048, and 8192, similar to what is done by Liu \emph{et al.} \cite{liu2016}. Therefore, the input temporal resolution is $16000/512=31.25$ frame-per-second (FPS) in the input feature maps. After the processing of the sound model with 16 total strides, the output has a temporal resolution of $31.25/16=1.95$ FPS. We also use this resolution as the temporal resolution for the action and object models. The log mel-spectrograms are extracted with Librosa, an open-source Python library for audio analysis \cite{librosa2017}. 
All the images and videos are resized so that the longer side has 256 pixels, maintaining the aspect ratio.

The RGB images are used in the object model. They are sampled from a video clip with 1.95 FPS, which is the same as the temporal resolution of the sound model. Because the FCNs can handle input of arbitrary sizes, we do not have to pad the images.

The dense optical flows are used in the action model. We extract the dense optical flows also with 1.95 FPS. For each frame, we use a stack of five optical flows as the representation, including the dense optical flow of the frame itself and the dense optical flows of its four neighboring frames (two after and two before). Each dense optical flow is decomposed into $x$-direction flow and $y$-direction flow, so there are totally $5\times 2=10$ channels for the input. We will test five temporal resolutions for the extraction of dense optical flows in Section \ref{sec:resolution}. To extract the dense optical flows, we convert RGB images to the gray scale and then employ OpenCV\footnote{\url{http://opencv.org}}.

\subsection{Datasets} \label{sec:dataset}
We use five datasets in this paper. For training the action and object models, we use a subset of YouTube-8M\footnote{\url{https://research.google.com/YouTube-8M/}} \cite{abu2016}. For training the sound model, we use the AudioSet\footnote{https://research.google.com/audioset/} \cite{jort2017audioset}. For evaluating the action models, we manually annotate action key points in video clips from 135 videos of YouTube-8M. For evaluating the object model, we collect a set of instrument images from ImageNet. For evaluating the sound model, we use MedleyDB \cite{bittner2014}. A list of used datasets can be found in TABLE \ref{tab:dataset}. 
In this paper, we focus on the detection of nine instruments.
The properties of the instruments and the number of data in the datasets we use are presented in TABLE \ref{tab:instrument}.

\begin{table*}[htbp]
\caption{Datasets used in this paper and their data types, annotation types, and usages. YT8M-IPA represents the proposed YouTube-8M-Instrument-Playing-Action dataset.}
\label{tab:dataset}
\centering
\begin{tabular}{l|lll}
\hline
Dataset							& Data type		& Annotation type						& Usage	in this paper \\
\hline
YouTube-8M \cite{abu2016}		& YouTube video	& Video-level label						& Training of the sound, object, and action models \\
YT8M-IPA						& YouTube video	& Playing-action key point we annotate	& Evaluation of the action model	\\
AudioSet \cite{jort2017audioset}& YouTube video	& Video-level sound label				& Training of the sound model		\\
MedleyDB \cite{bittner2014}		& Music audio	& Frame-level sound label				& Evaluation of the sound model		\\
ImageNet-Instrument\footnotemark		& Image			& Instrument bounding box				& Evaluation of the object model	\\
MagnaTagATune \cite{law2009magna}	& Music audio	& Track-level music-related label		& Training of the sound model		\\
\hline
\end{tabular}
\end{table*}

\footnotetext{http://www.image-net.org/}

\begin{table*}[htbp]
\caption{Models used in this paper and their input features, prediction types, and training targets.}
\label{tab:modellist}
\centering
\begin{tabular}{l|lll}
\hline
Model name							& Input feature 		& Prediction type	& Training target		\\
\hline
Sound model							& Log mel-spectrogram	& Sound activation	& Video-level instrument label \\
Object model						& RGB image				& Object activation	& Video-level instrument label \\
Video tag as target (VT)			& Dense optical flow	& Action activation	& Video-level instrument label \\
Sound as target (ST)				& Dense optical flow	& Action activation	& Sound activation \\
Object as target (OT)				& Dense optical flow	& Action activation	& Object activation \\
Sound$\times$Object as target (SOT)	& Dense optical flow	& Action activation	& Sound activation$\times$Object activation\\
\hline
\end{tabular}
\end{table*}


\subsubsection{YouTube-8M}

We use a subset of YouTube-8M dataset \cite{abu2016}. We collect videos for nine instruments according to the tag information provided by YouTube-8M. The nine instruments are `Accordion', `Cello', `Drummer', `Flute', `Guitar', `Piano', `Saxophone', `Trumpet', and `Violin'. \footnote{While there are certainly other instruments, we choose these nine instruments mainly for they cover instrument types that are commonly seen. On one hand, we have sufficient number of training data for each of them in the datasets we use. On the other hand, we still need to manually annotate the action locations of the chosen instruments for evaluation (because such labels are not available elsewhere) so we have to limit the number of instruments. As the proposed methodology is quite generic, we believe our model can be easily extended to deal with other instruments in the future work.}

Note that we choose `Drummer' instead `Drum' because we notice that the `Drummer' tag seems to contain more instrument-playing videos. We will refer to `Drummer' tag as `Drum' in the rest of this work. In addition, we find that the videos labeled with `Trumpet' in YouTube-8M contain not only videos of trumpets, but also videos of other instruments in the brass family, such as cornet, French horn, and trombone Therefore, we will treat the `Trumpet' as a more general trumpet-like tag.

YouTube-8M has divided the data into `train,' `validate,' and `test' sets. We collect 16,804 videos as the training set from YouTube-8M `train' set, and 2,100 videos as the validation set from YouTube-8M `validate' set. We use the first minute in each video clip for training. Each instrument has at least 2,000 videos for training. Note that we only need clip-level labels for the training and validation sets.

\begin{table*}[htbp]
	\renewcommand{\arraystretch}{1.3}
	\setlength{\tabcolsep}{4pt}
	\caption{Properties of the nine instruments. The lower part of the table contains the number of data in the datasets. In YT8M-IPA, the playing actions are annotated at the intersections of the action regions and the playing tools as described in Section \ref{sec:yt8m_instrument}. `frs' represents frames. `imgs' represents images.}
	\label{tab:instrument}
	\centering
	\begin{tabular}{l|lllllllll}
		\hline
			& Accordion	&Cello		&Drum		& Flute	& Guitar	& Piano	& Saxophone	& Trumpet	& Violin 	\\
		\hline
		Action region		& Keys/body	& Strings	& Drum skins	& Holes/mouthpiece	&	Strings	& Keys	& Keys/mouthpiece	& Valves/mouthpiece	& Strings \\
		Playing tool		&  	Hands	&	Hand/bow	& Sticks	& Hands/mouth	& Hands	& Hands	& Hands/mouth	& Hands/mouth	& Hand/bow \\

		Portable?			& \checkmark	& \checkmark	& $\times$	& \checkmark	& \checkmark	& $\times$	& \checkmark	& \checkmark	& \checkmark	\\
		\hline
		YouTube-8M (clips)		& 2279		& 2260		& 2495		& 2264	& 3204	& 3678	& 2367	& 2240	& 3521 \\
		YT8M-IPA (clips/frs)	& 15/600	& 15/600	& 15/600	& 15/600	& 15/600	& 15/600	& 15/600	& 15/600	& 15/600 \\
		AudioSet (clips)		& 2658		& 4664		& 4866		& 4281	& 5624	& 5233	& 2966	& 3654	& 6553 \\
		MedleyDB (songs)	& 5 & 11	& 65	& 10	& 64	& 43	& 5	& 7	& 14 \\
		ImageNet-Inst. (imgs)	& 412	& 323	& 252	& 359	& 135	& 315	& 343	& 294	& 365	\\

		\hline
	\end{tabular}
\end{table*}



\subsubsection{YouTube-8M-Instrument-Playing-Action} \label{sec:yt8m_instrument}

There are no action annotations in YouTube-8M, so we manually annotate a set of video clips from YouTube-8M. The metadata and video IDs of the YouTube-8M `test' set are not available, so we choose the testing data from YouTube-8M `validate' set, not overlapping with our validation set. 15 videos are chosen for each instrument. We manually annotate frames in the 0 to 10 seconds and 30 to 40 seconds so that we can evaluate the performance of our model for action detection. With the temporal resolution 1.95 FPS, this comprises 5,400 snapshots. This set of annotations is used only for evaluating action models, not for training. We will refer to this subset with manual annotations as YouTube-8M-Instrument-Playing-Action, or YT8M-IPA for short.

\begin{figure}[!t]
	\centering
	\includegraphics[width=0.95\columnwidth]{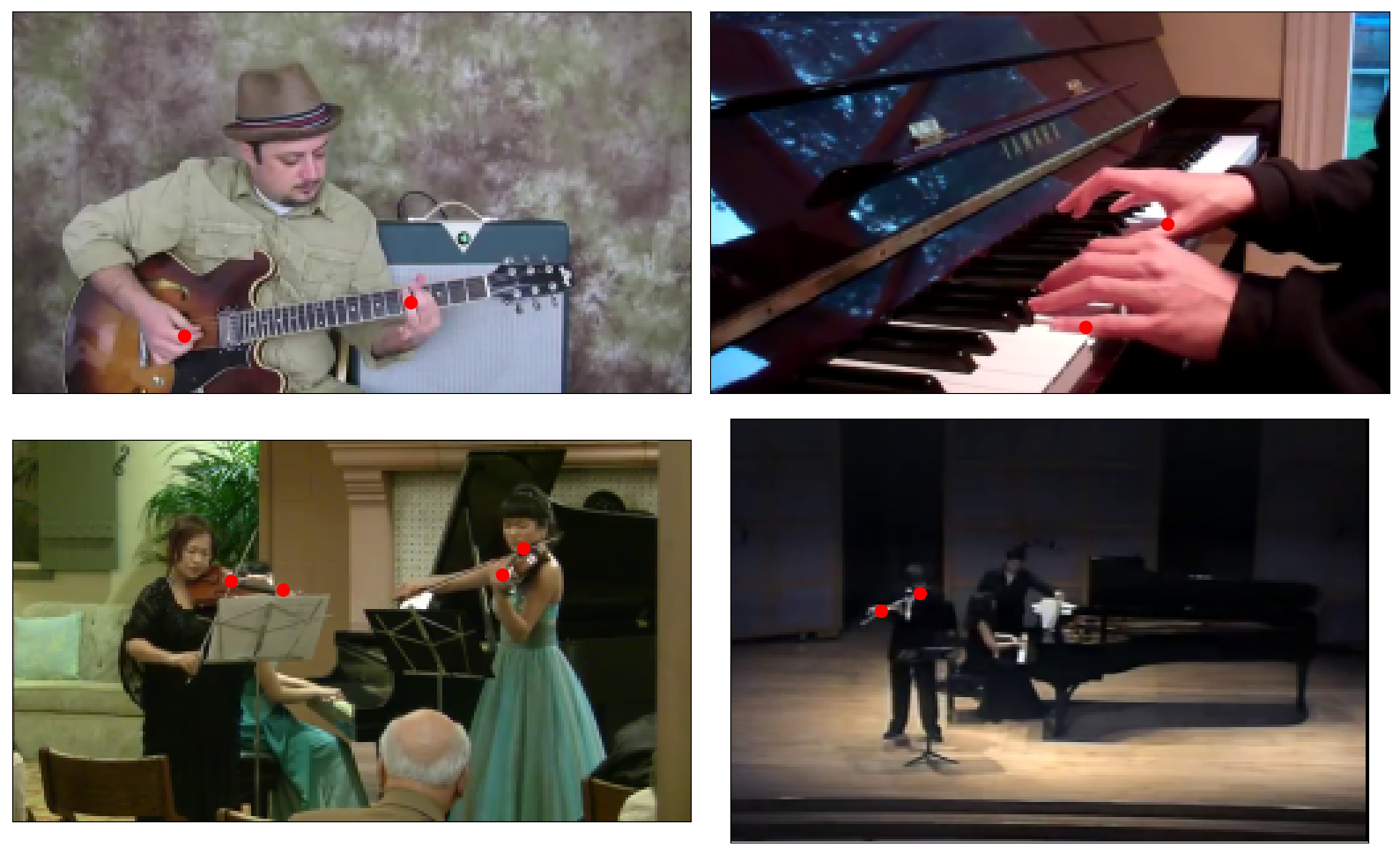}
	\caption{Examples of the manually annotated key points (red dots) of instrument-playing actions. We annotate the locations that are most directly responsible to making the instrument sounds as described in Section \ref{sec:yt8m_instrument}. Best seen in color. }
	\label{fig:anno}
\end{figure}

We represent the locations of instrument-playing actions as key points, instead of regions that are commonly used in the literature of action detection \cite{zhou2015, mosabbeb2015}. We choose to do so because we want to predict the actions that are most directly responsible for making instrument sounds. The sounds are usually made by the contacts between sound making tools, such as hands and sticks, and an instrument, and the contacts are usually more like points than regions. Some examples of the annotations are shown in Fig. \ref{fig:anno}.

The first author of the paper annotates the locations of the instrument playing according to the following principles. For wind instruments like flute, saxophone, and trumpet, the locations where the hands are pressing and the location of the mouth are labeled. For string instruments like cello, violin, and guitar, the location of the pressing hand and the intersection of the stroking hand (or the bow) and the strings are labeled. For accordion, the two hands are labeled, and the center of accordion is also labeled because the deformation of the accordion is also an indicator of playing. For drum and piano, the locations of the hands/sticks hitting the instruments are labeled. Note that these locations are labeled only if the instruments in sight are responsible for making the sounds at a given frame. 


\begin{table*}[htbp]
\renewcommand{\arraystretch}{1.3}
\setlength{\tabcolsep}{0.65em}
\caption{Evaluation of the sound models for instrument sound detection. The sound model trained with AudioSet outperforms the one trained with YouTube-8M and the one (the model in our previous work \cite{liu2016}) trained with the music dataset MagnaTagATune. We use `Acc.' as shorthand for Accordion.}
\label{tab:sound}
\centering
\begin{tabular}{l||ccccccccc||cc}
\hline
\multirow{2}{*}{Training data}		& Accordion	&Cello		&Drum		& Flute	& Guitar	& Piano	& Saxophone	& Trumpet	& Violin	& \multirow{2}{*}{AVG} & \multirow{2}{*}{AVG w/o Acc.} \\
	& sound	& sound	& sound	& sound	& sound	& sound	& sound	& sound	& sound	& & \\

\hline
MagnaTagATune  (MTT)	& NA		& 0.877	& 0.805	& 0.752	& 0.765	& 0.688	& 0.717	& 0.644	& 0.871	& NA		& 0.765 \\
Subset of YouTube-8M (YT8M)		& 0.806	& 0.805	& 0.705	& 0.803	& 0.740	& 0.779	& 0.747	& 0.911	& 0.878	& 0.797	& 0.796 \\
Subset of AudioSet (AS)	& 0.714	& 0.798	& 0.773	& 0.790	& 0.735	& 0.750	& 0.910	& 0.920	& 0.821	& \textbf{0.801}	& \textbf{0.812} \\
\hline
MTT+YT8M+AS & 0.724	& 0.871	& 0.763	& 0.802	& 0.771	& 0.761	& 0.860	& 0.886	& 0.855	& 0.810	& 0.821 \\

\hline

\end{tabular}
\end{table*}

\subsubsection{ImageNet-Instrument} \label{sec:imagenet_instrument}
In order to evaluate the object localization ability of the object model, we collect images of the nine instruments from the ImageNet website. They also provide the bounding boxes for the locations of the instruments. We collect totally 2,798 images and the corresponding bounding boxes. An instrument has on average 311 images ranging from 135 to 412.

\subsubsection{MedleyDB} 
MedleyDB \cite{bittner2014} is a multi-track instrument dataset. It also contains the timestamps of the occurrences of the instrument sounds. There are totally 111 songs with the nine instruments used in this work.\footnote{We aggregate the `acoustic guitar,' `clean electric guitar,' and `distorted electric guitar' in MedleyDB into the `Guitar' tag, and aggregate `baritone saxophone,' `soprano saxophone,' and `tenor saxophone' in MedleyDB into the `Saxophone' tag.} We use it to evaluate the sound model.

\subsubsection{AudioSet}

AudioSet \cite{jort2017audioset} is a video dataset released by Google, containing audio annotations on a 10-second clip in each of the videos. We collect a subset of the nine instruments from AudioSet, consisting of 35,512 video clips for training and 902 video clips for validation. Each instrument has 3,945 training clips on average.

\subsection{Training}
For training the action and object models, we use the first 60 seconds in each training video clip from YouTube-8M as the training data. The 60-second video clip is divided into 12 consecutive 5-second sub-clips. Under the resolution of 1.95 FPS, each sub-video contains 10 frames. For each mini-batch, a video is randomly picked without replacement and then a sub-clip is randomly picked from the 12 sub-clips of the video, so the size of a mini-batch is 10. We follow Simonyan \emph{et al.} \cite{simonyan2014} to use stochastic gradient descent with 0.9 momentum, but fix the learning rate to 0.001. For training the sound model, the 10-second annotated sub-clip from AudioSet is used, and we follow Liu \emph{et al.} \cite{liu2016} to use AdaGrad \cite{duchi2011} with 0.01 initial learning rate. The loss function for all models is the binary cross-entropy, $-(q\log(p)+(1-q)\log(1-p))$, where $q$ is the target and $p$ is the output of a model.

We find that directly training the object model with pre-trained parameters would not perform well, even worse than training from scratch without pre-trained parameters. The loss almost never decreases. This could be due to the mismatch of scales in the pre-trained parameters in the early convolutions and the randomly sampled parameters in the late convolutions, which makes the back-propagation difficult. Therefore, we adopt a two-step training process. First we freeze the parameters in the early convolutions and only allow updating the parameters in the late convolutions. After 20 epochs of training, we free the parameters in the early convolutions so that they start to update for 30 more epochs. For the sound model, 100 epochs are executed.
For the action models, 100 epochs are executed for the experiments in Section \ref{sec:resolution}. The remaining experiments thereafter will use the best one in Section \ref{sec:resolution} as a pre-trained model and train for another 30 epochs. The parameters from the epoch with the best clip-level AUC\footnote{We do not use frame-level AUC here for we do not have frame-level annotations at all for the training and validation sets. AUC stands for Area Under the Curve, a widely used metric for classification problems.} are picked as the parameters for testing. We implement our models with PyTorch\footnote{\url{http://pytorch.org}}.

\section{Experiments} \label{sec:exp}

We report our experiments in this section. We will first evaluate independently the performance of the sound model (for instrument sound detection) and the object model (for instrument object detection). In the evaluation of the action model (for playing action detection), we will first see the effect of the temporal resolution of dense optical flows, and then we evaluate the performance of the action models trained with the four different targets discussed in Section \ref{sec:method}.

There are three models interact with each other, so we intend to investigate the performance of each model and also how the two auxiliary models affect the performance of the action model. Specifically, we want to answer the following questions:
\begin{enumerate}
\item How does the resolution of dense optical flows affect the performance of playing action detection?
\item Do the supervisions provided by the sound and object models improve the performance of playing action detection?
\item How does the effectiveness of the sound model affect the effectiveness of the ST and SOT action models?
\item Does the fusion with different streams of modalities (sound and object) improve the performance of playing action detection?
\end{enumerate}

\subsection{Performance of the sound model for instrument sound detection}

We evaluate the ability of temporal localization of the sound model with the music dataset MedleyDB. We compute the per-class AUC by taking each frame as an instance and computing the score over all frames.

\begin{table*}[htbp]
\renewcommand{\arraystretch}{1.3}
\caption{Evaluation of the object models for instrument object detection. They are tested on the instrument images with bounding boxes collected from ImageNet (HIT RATE). The one initialized with a pre-trained model VGG\_CNN\_M\_2048 (With pre-training) outperforms the one with random initialization (From scratch) by a large margin.}
\label{tab:object}
\centering
\begin{tabular}{l||ccccccccc||c}
\hline
			& Accordion	&Cello		&Drum		& Flute	& Guitar	& Piano	& Saxophone	& Trumpet	& Violin	& \multirow{2}{*}{Average}	\\
			& object	& object	& object	& object	& object	& object	& object	& object	& object	& \\
\hline
From scratch		& 0.845	& 0.740	& 0.540	& 0.345	& 0.719	& 0.841	& 0.770	& 0.507	& 0.627	& 0.659 \\
With pre-training	& 0.927	& 0.916	& 0.579	& 0.671	& 0.963	& 0.946	& 0.945	& 0.833	& 0.847 & 0.848 \\
\hline
\end{tabular}
\end{table*}

We compare three sound models. The first one is trained with YouTube-8M and the architecture in TABLE \ref{tab:sound_structure}. The second is trained with AudioSet and the same architecture. The third model is the one presented in our previous work \cite{liu2016}\footnote{\url{https://github.com/ciaua/clip2frame/blob/master/data/models/model.20160309_111546.npz}}, which was trained with the music dataset MagnaTagATune \cite{law2009magna}. The output of the third model covers eight out of the nine instruments used in this paper except for accordion, so we compute the AUCs of these eight instruments for this model. The result is shown in TABLE \ref{tab:sound}.

In general, when the model is trained with a single dataset, the best result is obtained by the model trained with AudioSet, achieving 0.801 AUC, and 0.812 AUC excluding the Accordion sound. This can be expected because AudioSet has better annotations. The YouTube-8M sound model is close to the AudioSet sound model, achieving 0.797 AUC (0.796 AUC excluding the Accordion sound). Although the MagnaTagATune sound model is in general inferior to the other two models, it performs well in some instrument sounds, such as the Cello sound, the Drum sound, and the Violin sound. 

We also train an additional model with the data combining MagnaTagATune, YouTube-8M, and AudioSet. It outperforms the models trained with a single dataset, achieving 0.810 AUC, and 0.821 AUC excluding Accordion.

Although the model trained with the data combining the three datasets achieves the best performance when it is evaluated with MedleyDB, we will still mainly use the AudioSet sound model as the sound model in the following experiments for two reasons. 
First, we find that ST action models trained with an AudioSet sound model will yield better performance than those trained with an MTT+YT8M+AudioSet sound model, as we will show in Section \ref{sec:compare_sound_models}.
Second, we conceive it is better to use only one dataset for training, if possible, for the sake of simplicity. In addition, as the AudioSet actually covers many more instruments and other action types, in the future one can use the same data source for extension.

\subsection{Performance of the object model for instrument object detection} \label{sec:object_eval}

We evaluate the ability of spatial localization of the object model independently with ImageNet instrument images. ImageNet website provides the bounding boxes of the locations of the instruments. We compute the accuracy in a way similar to Oquab \emph{et al.} \cite{oquab2015}. Each instrument is processed separately. For an image in an instrument class, the prediction of the object is considered as a hit if the location of the maximum prediction value is inside the bounding box. 

The result is shown in Table \ref{tab:object}. The model with pre-training achieves $0.848 \%$ on average, which is greatly higher than the $0.659 \%$ obtained by the model trained from scratch. We refer to the model with pre-training as the object model hereafter. Most instruments have hit rate higher than $80\%$. The Flute object is one of the most challenging cases due to its thin body. The Drum object achieves the worst hit rate, 0.579. By looking into the images collected from ImageNet, we find that the images include various types of drums. As far as we have seen, most videos with `Drummer' tag in YouTube-8M have jazz drum set. Therefore, the object model trained with them could perform worse for other types of drums.

\begin{figure}[!t]
\centering
\includegraphics[width=0.9\columnwidth]{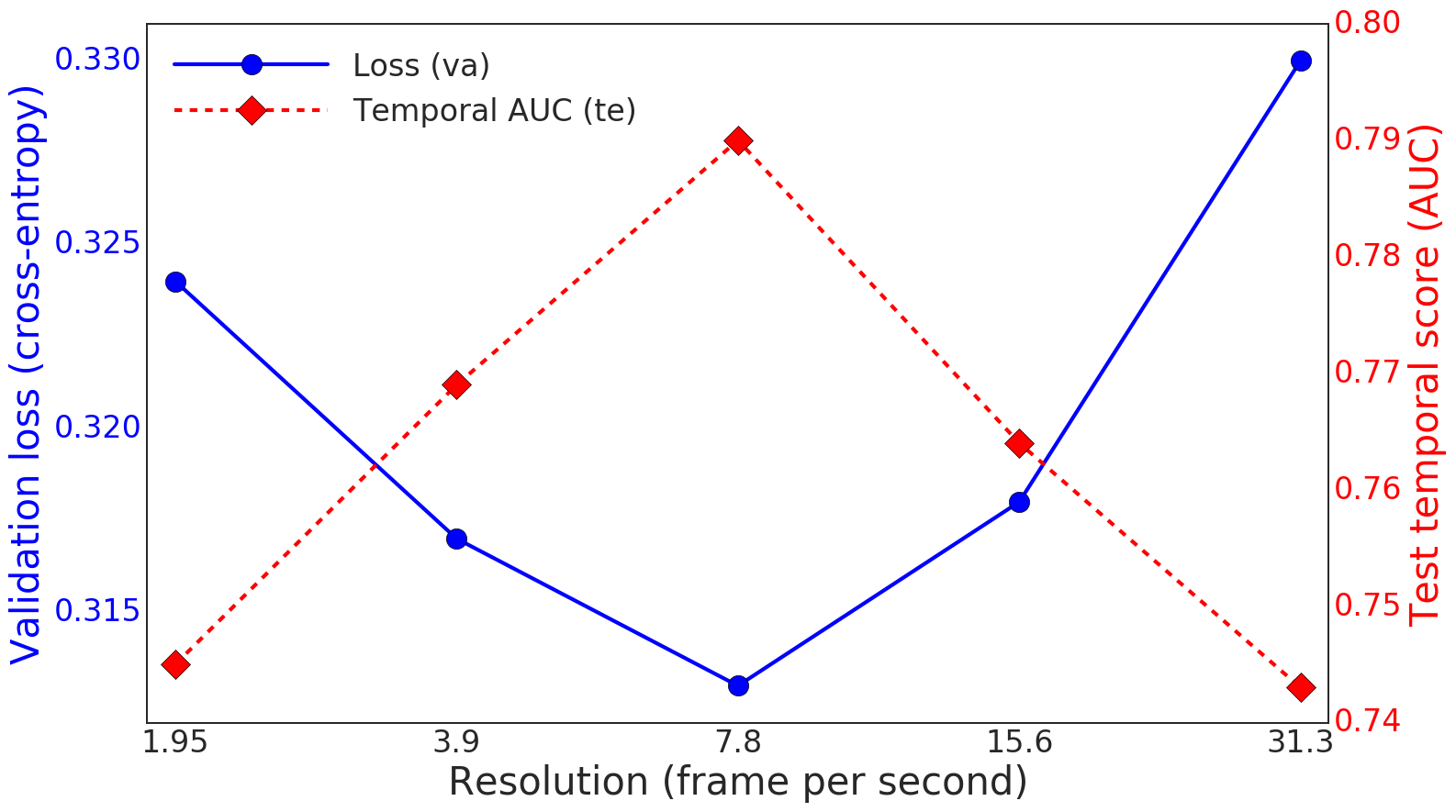}
\caption{Effect of different temporal resolutions of dense optical flows. The blue solid line and the red dashed line are the validation losses and the test temporal AUC, respectively, of the action models trained with different resolutions of dense optical flows. The validation loss (the lower the better) and the test AUC (the higher the better) are both optimal at the resolution of 7.8 frame-per-second.}
\label{fig:resolution}
\end{figure}

\subsection{Performance of the action model for instrument-playing action detection}
We evaluate the temporal and spatial accuracy of the proposed models. We have manually constructed the test set by annotating the locations that are relevant to the playing actions, as mentioned in Section \ref{sec:setup}. For the temporal evaluation, a frame is a positive instance if the set of annotated locations of the frame is non-empty, and a negative instance otherwise. We evaluate each instrument separately with AUC. The AUC is first computed on the frames within a video clip and then the scores are averaged over all the video clips.


For spatial evaluation, the evaluation is performed in a way similar to the evaluation of the object model by evaluating the location of the maximum prediction value in a frame, referred to as the \emph{max location}. However, the action locations are represented as coordinates instead of bounding boxes, so we compute the minimum distance between the max location and the annotated locations, that is, 
\begin{equation}
D=\min_{l\in L} \lVert l-M \rVert_2,
\end{equation}
where $L$ is the set of the coordinates of the annotated locations in a frame, and $M$ is the coordinate of the max location.\footnote{
We note that intersection-over-union (IoU) is a metric that is commonly used in action detection or object detection that involves prediction of regions \cite{zhou2015, mosabbeb2015, shou2016, ren2015, girshick2015}, in which actions are attributed to the entire object in action, such as a human being or an animal. We opt for not using IoU in our work, because the locations of the instrument-playing actions we consider in this paper are composed of very small parts and, hence, it is more suitable to represent them as key points, as described in Section \ref{sec:yt8m_instrument}. \color{black}
}

\subsubsection{Effect of resolutions of dense optical flows} \label{sec:resolution}

The temporal resolution will affect how subtle the movements the model can see, so we investigate the choice of the temporal resolution of dense optical flows for the action model. We use a stack of 5 consecutive dense optical flows. We test five resolutions that are multiples of the base temporal resolution, $1.95$, $3.9$, $7.8$, $15.6$, and $31.3$ FPS. The last one is roughly the default FPS of most online videos.

\begin{table}[htbp]
\renewcommand{\arraystretch}{1.3}
\caption{Evaluation of the action models for instrument-playing action detection. The ST and SOT models outperform other models temporally because they have temporal supervisions from the sound model, while the OT and SOT models outperform other models spatially because they have spatial supervisions from the object model.}
\label{tab:action_average}
\centering
\begin{tabular}{c||cc||cc}
\hline
					& \multicolumn{2}{c||}{Threshold}		& \multicolumn{2}{c}{Score}		\\
\cline{2-5}
Model					& \multirow{2}{*}{Sound}	& \multirow{2}{*}{Object} 		& Temporal 		& Spatial 	\\
					&			&			& (AUC)		& (pixel)	\\
\hline
Center 				& NA			& NA			& NA			& 37.2				 \\
Sound model				& NA			& NA			& 0.881		& NA				 \\
Object model				& NA			& NA			& 0.712		& 33.3				 \\
\hline
Video tag as target (VT)		& NA			& NA			& 0.790		& 33.9				 \\
\hline
Sound as target (ST01)		& 0.1			& NA			& 0.824		& 32.7				 \\
Sound as target (ST03)		& 0.3			& NA			& 0.827		& 32.2				 \\
Sound as target (ST05)		& 0.5			& NA			& 0.827		& 33.7				 \\
Sound as target (ST07)		& 0.7			& NA			& 0.830		& 33.0				 \\
Sound as target (ST09)		& 0.9			& NA			& 0.824		& 33.7				 \\
\hline
Object as target (OT01)		& NA			& 0.1			& 0.680		& 45.8				 \\
Object as target (OT03)		& NA			& 0.3			& 0.809		& 29.4				 \\
Object as target (OT05)		& NA			& 0.5			& 0.809		& 29.5				 \\
Object as target (OT07)		& NA			& 0.7			& 0.798		& 29.2				 \\
Object as target (OT09)		& NA			& 0.9			& 0.788		& 31.0				 \\
\hline
S$\times$O as target (SOT0501) 	& 0.5			& 0.1			& 0.828		& 29.4				 \\
S$\times$O as target (SOT0503) 	& 0.5			& 0.3			& \textbf{0.834}	& 29.1				 \\
S$\times$O as target (SOT0505) 	& 0.5			& 0.5			& 0.827		& \textbf{28.9}		 \\
S$\times$O as target (SOT0507) 	& 0.5			& 0.7			& 0.820		& 29.5				 \\
S$\times$O as target (SOT0509) 	& 0.5			& 0.9			& 0.803		& 30.7				 \\
\hline
\end{tabular}
\end{table}

\begin{table*}[htbp]
\renewcommand{\arraystretch}{1.3}
\caption{Instrument-wise action detection performance. The best temporal scores are all achieved by ST and SOT that have temporal supervisions from the sound model. Similarly, the best spatial scores are mostly achieved by OT and SOT that have spatial supervisions from the object model. }
\label{tab:action_classwise}
\centering
\begin{tabular}{cc||ccccccccc||c}
\hline
				& \multirow{2}{*}{Model (thresholds)}		& Accordion	&Cello		&Drum		& Flute	& Guitar	& Piano	& Saxophone	& Trumpet	& Violin & \multirow{2}{*}{Average} \\
				& & action	& action	& action	& action	& action	& action	& action	& action	& action	\\
\hline
				& Object model			& 0.695	& 0.812	& 0.609	& 0.751	& 0.723	& 0.582	& 0.715	& 0.694	& 0.823	& 0.712 \\
				& Video tag as target		& 0.856	& 0.918	& 0.824	& 0.729	& 0.813	& 0.742	& 0.666	& 0.685	& 0.875 & 0.790 \\
Temporal			& Sound as target (0.5)		& 0.866	& 0.919	& \textbf{0.853}	& 0.783	& \textbf{0.871}	& 0.770	& 0.741	& 0.757	& 0.881 & 0.827 \\
(AUC)				& Object as target (0.3)		& 0.887	& 0.932	& 0.823	& 0.744	& 0.787	& 0.697	& 0.751	& 0.772	& 0.891 & 0.809 \\
				& S$\times$O as target (0.5, 0.3)	& \textbf{0.894}	& \textbf{0.937}	& 0.846	& \textbf{0.785}	& 0.838	& \textbf{0.778}	& \textbf{0.754}	& \textbf{0.777}	& \textbf{0.899} & \textbf{0.834} \\
\hline
				& Object model			& 27.7		& \textbf{21.8}		& 56.7		& 27.6		& 29.5		& 51.1		& 25.3		& 39.6		& 20.5	 & 33.3 \\
				& Video tag as target		& 24.9		& 28.8		& 42.2		& 30.5		& \textbf{28.1}		& \textbf{31.7}		& 38.3		& 56.7		& 23.9	 & 33.9 \\
Spatial			& Sound as target (0.5)		& 27.1		& 27.2		& 41.8		& 36.5		& 29.3		& 32.9		& 30.5		& 49.8		& 24.5	 & 33.3 \\
(pixel)				& Object as target (0.3)		& 24.3		& 23.1		& 45.1		& 27.4		& 29.2		& 39.4		& \textbf{23.2}		& \textbf{34.3}		& \textbf{19.0}	 & 29.4 \\
				& S$\times$O as target (0.5, 0.3)	& \textbf{23.8}		& 25.9		& \textbf{40.7}		& \textbf{26.9}		& 28.4		& 31.9		& 26.6		& 38.1		& 19.7	 & \textbf{29.1} \\

\hline
\end{tabular}
\end{table*}

Let's assume for a moment that we are using a resolution of 2 FPS. A stack of 5 frames cover about $5\times(1/2)=2.5$ second. If we use a finer resolution, say 8 FPS, the stack will cover $5\times(1/8)=0.625$ second. The movements of instrument-playing actions are not always fast, so we will desire the feature to cover a reasonably span of time that is long enough to capture the playing movements. The simplest way to cover a larger period of time is to use a larger stack of dense optical flows, say 20 consecutive frames, but this will also largely increase the computational time and resource. To keep the experiments manageable, we will maintain a stack of 5 frames and only vary the temporal resolutions.

We measure the effectiveness with the validation loss and also the temporal accuracy on the test set. The result is shown in Fig. \ref{fig:resolution}. We can see that both measures favor the resolution of 7.8 FPS. Therefore, we will use this resolution in all the experiments that follow. A stack of 5 consecutive dense optical flows with this resolution will cover 0.64 second.

\subsubsection{Effect of different training targets} 

In this subsection, we evaluate the performance of action models trained with four different targets discussed in Section \ref{sec:method}. In addition to the models trained with the four types of targets, we also include three baseline models. The first one is the object model for both the temporal and the spatial evaluations, that is, predicting the presence of playing actions simply by the presence of the instruments. The second one is the sound model for the temporal evaluation, that is, predicting the presence of playing actions simply by the presence of the instrument sounds. The third one is for the spatial evaluation by always predicting the center of the scene in each frame. It is sometimes a good guess for videos because the cameras are often centered at the player of the main instrument.

We show the average result over all nine instruments in TABLE \ref{tab:action_average}. The sound model performs very well temporally. It verifies that the sounds are indeed important cues for the playing actions. The object model performs well spatially because the characterizing portion of an instrument is often also the sounding part. Predicting the center is among the worst models.

We will use abbreviations of the format ``$<$training target$>$$<$threshold $v$ for the sound model if available$>$$<$threshold $u$ for the object model if available$>$'' as the name for an action model. For example, OT03 is the action model trained with the target produced by the object model with a threshold $u=0.3$, and SOT0503 is the action model trained with the target produced by the sound model with a threshold $v=0.5$ and the object model with a threshold $u=0.3$. In general, we can see that the ST models outperform the VT model temporally, OT models outperform the VT model spatially, and the SOT models outperforms the VT model both temporally and spatially, when the thresholds are not too extreme, that is, between 0.3 and 0.7.

\begin{figure*}[!t]
\centering
\includegraphics[width=0.9\textwidth]{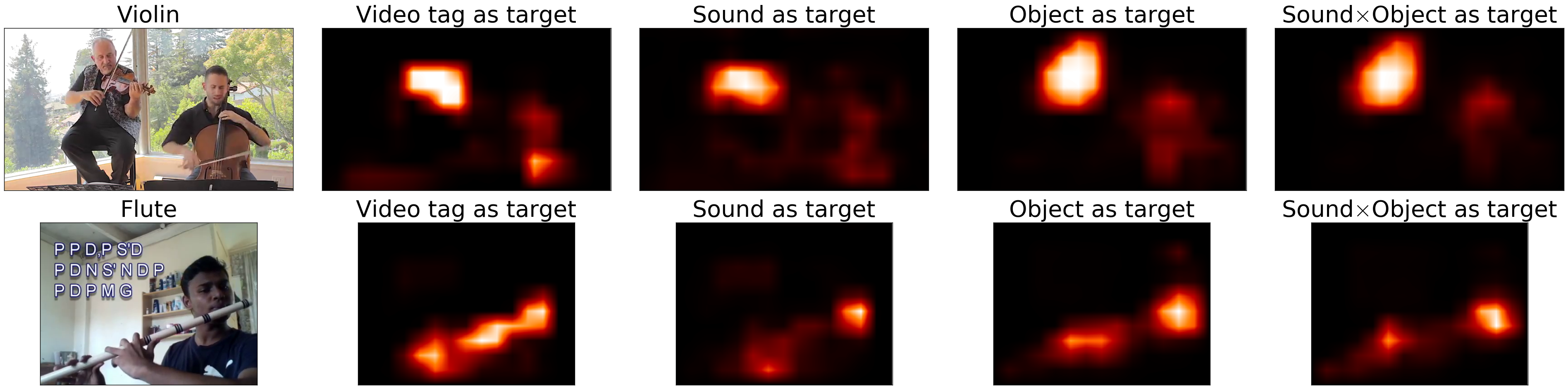}
\caption{The result of instrument-playing action detection as a function of different training targets. As the level of supervision changes from left to right, the result of action detection becomes cleaner and more accurate. The original videos of the two examples are uploaded by Jeremy Cohen (YouTue ID: 2G2VaBX24So) and Krishan Chotoe (YouTube ID: 55\_RhFOyRgk), respectively, both of which are under Creative Common license.}
\label{fig:comparison}
\end{figure*}

\begin{table*}[htbp]
\renewcommand{\arraystretch}{1.1}
\caption{Temporal performance of the ST action models by using as the training target either the sound model trained with the video dataset AudioSet or the one trained with the music audio dataset MagnaTagATune. The ST action model using the AudioSet sound model outperforms the one using the MagnaTagATune sound model in all instruments.}
\label{tab:sound_as_target}
\centering
\begin{tabular}{l||ccccccccc||cc}
\hline
\multirow{2}{*}{Action model (sound threshold)}		& Accordion	&Cello		&Drum		& Flute	& Guitar	& Piano	& Saxophone	& Trumpet	& Violin	& \multirow{2}{*}{Average w/o Acc.} \\
	& action	& action	& action	& action	& action	& action	& action	& action	& action	&  \\
\hline
ST: AudioSet	 sound (0.5)			& 0.866	& \textbf{0.919}	& \textbf{0.853}	& \textbf{0.783}	& 0.871	& \textbf{0.770}	& 0.741	& \textbf{0.757}	& \textbf{0.881}	& \textbf{0.822}  \\
ST: MagnaTagATune sound (0.5)	& NA		& 0.813	& 0.751	& 0.565	& 0.718	& 0.522	& 0.644	& 0.547	& 0.799	& 0.670  \\
ST: MagnaTagATune sound (0.3)	& NA		& 0.855	& 0.710	& 0.588	& 0.676	& 0.498	& 0.558	& 0.522	& 0.777	& 0.648  \\
ST: MagnaTagATune sound (0.1)	& NA		& 0.825	& 0.822	& 0.583	& 0.719	& 0.462	& 0.535	& 0.529	& 0.792	& 0.658  \\
\hline
ST: MTT+YT8M+AS sound (0.5)		& \textbf{0.873}		& 0.911 & 0.847 & 0.764 & \textbf{0.874} & 0.799	& \textbf{0.748} & 0.704	& 0.872 & 0.815	\\
\hline
\end{tabular}

\end{table*}

We also perform t-test on some pairs of models with 0.05 confidence level over the averages of all test videos. We found that ST05 is significantly better than VT and all OT models temporally and OT03 is significantly better than VT and all ST models spatially. Similarly, SOT0503 is significantly better than those without temporal (or spatial) supervision temporally (or spatially). The ST models are provided with temporal supervision so the performance is improved most temporally over the VT baseline model, while has less temporal improvement. On the other hand, the OT models are provided with spatial supervision so the performance is improved most spatially over the VT baseline model while have less spatial improvement. The SOT models are provided with both temporal and spatial supervisions so the performance is improved both temporally and spatially, compared with the VT model. The best temporal score 0.834 AUC and the best spatial score 28.9 pixels are both achieved by the SOT models. Thresholds have some impact on the performance for either ST, OT, or SOT models, but the performance is pretty stable if we choose thresholds around 0.5.

Next, we show the instrument-wise result in TABLE \ref{tab:action_classwise}. 
The best temporal scores are all achieved by ST and SOT that have temporal supervisions from the sound model. Similarly, the best spatial scores are mostly achieved by OT and SOT that have spatial supervisions from the object model. The instruments in wind family are among the most challenging cases. The object model tends to fail for actions of instruments that are difficult for the player to carry along. For example, we can see that the object model performs poorly in detecting the Drum action and the Piano action, achieving 0.609 and 0.582 AUCs, respectively. For the temporal performance, the inclusion of temporal supervision (ST and SOT models) improves the result for action detection of all the instruments, especially for the instruments in the wind family, Flute, Saxophone, and Trumpet. The movements of the instruments in the wind family are subtle, so the difference of positive and negative frames is small without temporal supervision. For the spatial performance, the object model again performs poorly in detecting the Drum action and the Piano action due to the large sizes of them. The action detections of the wind family instruments Flute, Saxophone, and Trumpet have large gains from the spatial supervision provided by the object model. Cello and Violin actions are also improved.

In general, the supervisions provided by the two auxiliary models significantly improve the performance temporally and spatially. We can also see this visually in Fig. \ref{fig:comparison}. Comparing to the result of the other three models, the result of SOT is more concentrated and less noisy.



\subsubsection{Effect of different sound models} \label{sec:compare_sound_models}

In this subsection, we want to see how the performance of the sound model affects the performance of the ST models for action detection. We compare the ST model trained with the AudioSet sound model and the ST model trained with the MagnaTagATune sound model. 
This comparison is interesting because there are performance gaps between the two models for some instruments in instrument sound detection, as shown in TABLE \ref{tab:sound}. Additionally, we also train an ST model with the MTT+YT8M+AS sound model.
The result of this experiment is shown in TABLE \ref{tab:sound_as_target}.

First, we observe that the ST models trained with the MTT+YT8M+AS sound model and the AudioSet sound model have close performance. Accordingly, we only consider the AudioSet sound model hereafter for its simplicity. 

From the result shown in TABLE \ref{tab:sound}, we can see that the MagnaTagATune sound model itself performs poorly in the Piano sound, the Saxophone sound, and the Trumpet sound, so it is not surprising that the ST action models trained with MagnaTagATune sound model also perform poorly in the Piano action, the Saxophone action, and the Trumpet action. 

\begin{table*}[htbp]
\renewcommand{\arraystretch}{1}
\caption{Fusion of different streams of modalities after training. The output of an action model is fused with the output of the sound model and/or the output of the object model by point-wise multiplication. The fusion significantly improves the result.}
\label{tab:fusion}
\centering
\begin{tabular}{cc||ccccccccc||c}
\hline
				&					& Accordion	&Cello		&Drum		& Flute	& Guitar	& Piano	& Saxophone	& Trumpet	& Violin & \multirow{2}{*}{Average} \\
				&					& action	& action	& action	& action	& action	& action	& action	& action	& action	&	\\
\hline
	& SOT (0.5, 0.3)	& 0.894	& 0.937	& 0.846	& 0.785	& 0.838	& 0.778	& 0.754	& 0.777	& 0.899 & 0.834 \\
				& SOT$\times$Object	& 0.895	& 0.945	& 0.847	& 0.799	& \textbf{0.875}	& 0.830	& 0.767	& 0.811	& 0.913 & 0.854 \\
Temporal		& SOT$\times$Sound	& 0.951	& 0.965	& \textbf{0.900}	& 0.912	& 0.850	& 0.932	& 0.884	& 0.888	& \textbf{0.931} & 0.912 \\
(AUC)				& SOT$\times$Object$\times$Sound	& \textbf{0.954}	& \textbf{0.974}	& 0.895	& \textbf{0.921}	& 0.857	& \textbf{0.943}	& \textbf{0.888}	& \textbf{0.892} 	& 0.929 & \textbf{0.917} \\
				& VT$\times$Object$\times$Sound	& 0.948	& 0.965	& 0.892	& 0.919	& 0.846	& 0.942	& 0.878	& 0.878	& 0.919	& 0.910 \\
\hline
	& SOT (0.5, 0.3)	& \textbf{23.8}		& 25.9		& 40.7		& 26.9		& 28.4		& 31.9		& 26.6		& 38.1		& 19.7 & 29.1 \\
Spatial				& SOT$\times$Object	& \textbf{23.8}		& \textbf{20.2}		& \textbf{37.5}		& \textbf{22.5}		& \textbf{26.2}		& \textbf{27.8}		& \textbf{20.2}		& \textbf{31.0}		& \textbf{18.6}	& \textbf{25.3} \\
(Pixel)				& VT$\times$Object		& 24.1		& 20.4		& 39.2		& 24.6		& 27.1		& 31.5		& 34.0		& 43.9		& 18.9	&29.3 \\

\hline
\end{tabular}
\end{table*}

On the other hand, we can see that the MagnaTagATune sound model outperforms the AudioSet sound model in the Cello sound, the Drum sound, the Guitar sound, and the Violin sound when they are evaluated with MedleyDB in TABLE \ref{tab:sound}. However, the ST action models trained with the MagnaTagATune sound model are still inferior to the one trained with the AudioSet sound model in the actions of these instruments. The main reason could be the mismatch of the recording conditions. While the evaluation shown in TABLE \ref{tab:sound_as_target} may favor the ST model trained with the AudioSet sound model as the action test set (i.e. YouTube-8M) is composed of online videos, the evaluation shown in TABLE \ref{tab:sound} may have given the MagnaTagATune sound model some advantages as the sound test set (i.e. MedleyDB) is composed of also high-quality music audios.

In summary, we can see that the quality of the sound model has significant impact on the performance of the action models supervised by them.

\subsection{Fusion of different streams after training} \label{sec:fusion}

In this subsection, we want to see how much the inclusion of other modalities as input information helps the detection by using the fusion scheme described in Section \ref{sec:fusion_method}. The sound model and the object model are only used as the training target in the previous subsections, while the predictions of the sound model and the object model are directly used to assist the prediction of the action model in this subsection.

The fusion result is shown in TABLE \ref{tab:fusion}. The fusion significantly improves the original action detection result using only the action model. For the temporal performance, the SOT without fusion achieves 0.834 AUC on average, while SOT$\times$Object$\times$Sound improves to 0.917 AUC. For the spatial performance, the SOT without fusion achieves 29.1 pixels, while the SOT$\times$Object improves to 25.3 pixels. The action detections of the instruments in the wind family are again among those that obtain the largest gains. We can also see that SOT still outperforms VT after fusion, which again verifies the importance of providing better supervisions. The temporal results are much closer after fusion for SOT and VT, probably because the sound model dominates the fused prediction. The spatial result of SOT still outperforms VT by a large margin.


\begin{figure}[!t]
\centering
\includegraphics[width=0.85\columnwidth]{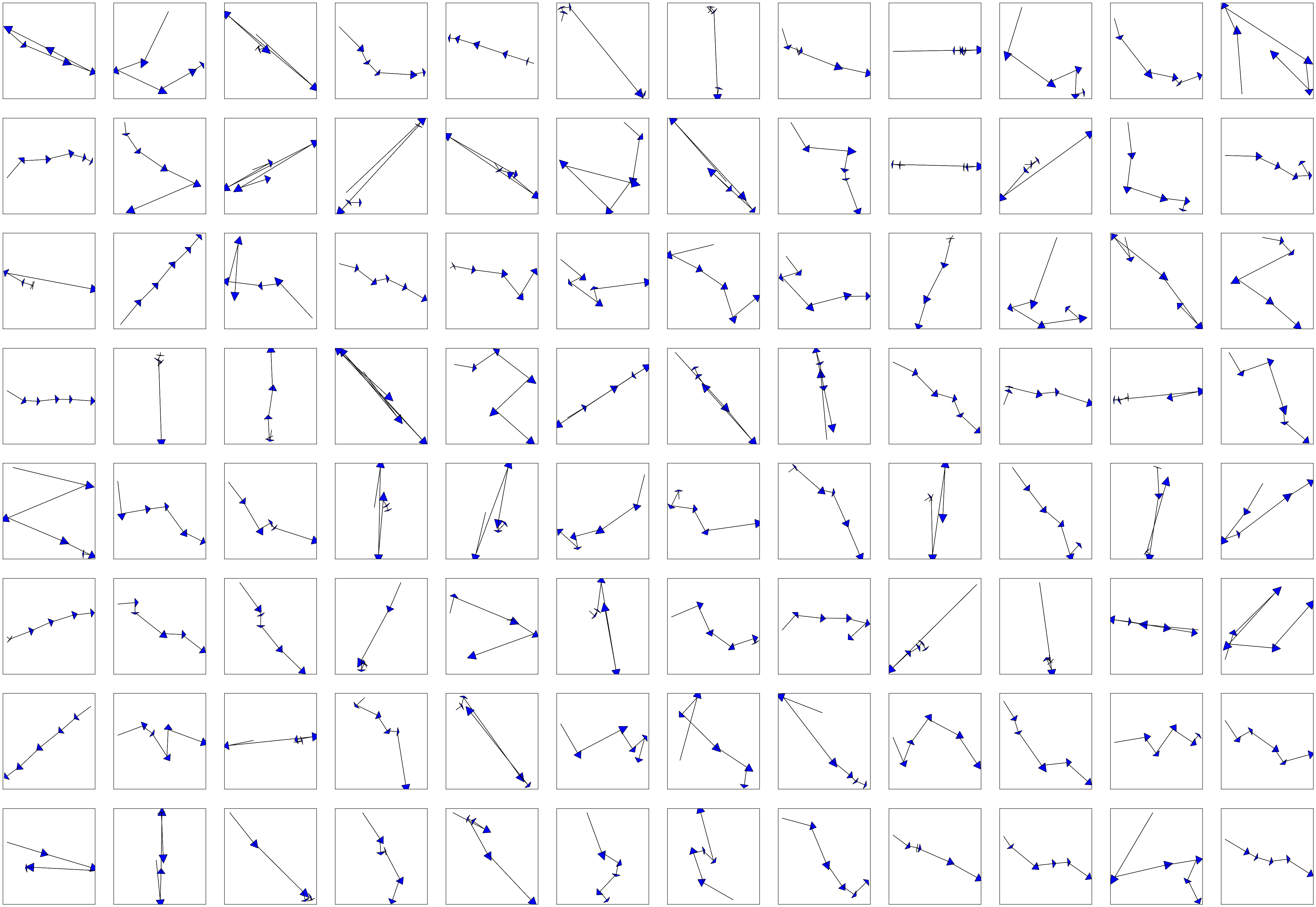}
\caption{Visualization of all 96 filters from Conv1 of the action model SOT0503. }
\label{fig:conv1}
\end{figure}

\subsection{Learned movements in the action model} \label{sec:filters}
We want to see what movements are learned in the action model in this subsection. Specifically, we visualize the learned filters in the first convolution layer, Conv1. The convolution layer, Conv1, is basically doing inner product between a filter, a $10\times7\times7$ tensor $[W_{k, i, j}]$ and a patch in the input feature map. For the action model, we take a stack of five $(x, y)$ coordinates at each location as the input feature, indicating the movements directions of five consecutive frames. Therefore, we can see a filter as indicating the important patterns in the input features. By averaging over the $7\times7$ receptive field in a filter, we get a $10$-dimensional vector, or, equivalently, a $5\times2$ matrix composed of five vectors. We show the 96 filters in Conv1 in Fig. \ref{fig:conv1} by plotting the five vectors one after the previous one to form a continuous movement. We can see that the network has learned different types of filters.

%
%

\begin{figure}[!t]
\centering
\includegraphics[width=0.9\columnwidth]{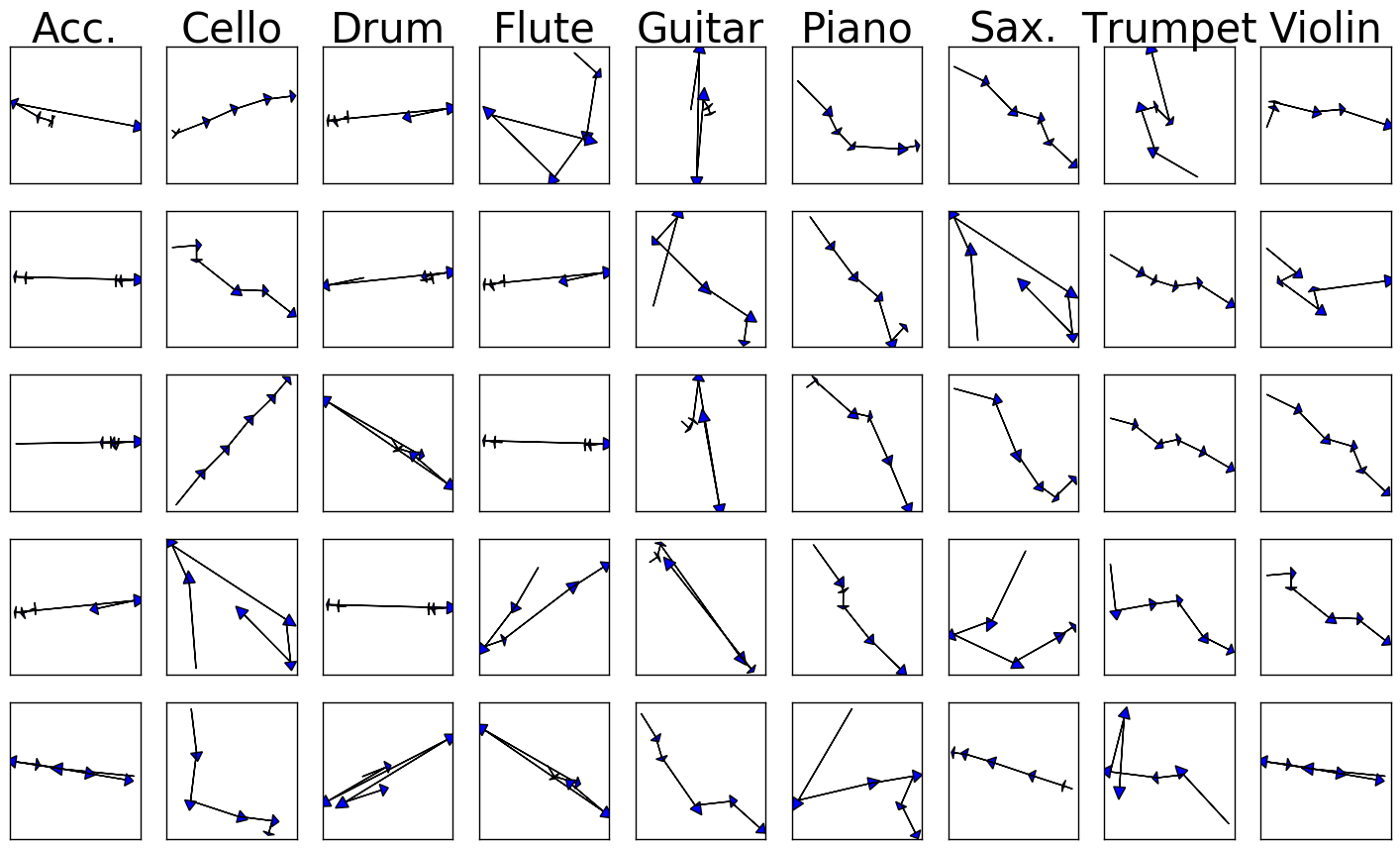}
\caption{Characterizing Conv1 filters for each instrument from the action model SOT0503. Each column is a subset of the 96 filters in Fig. \ref{fig:conv1}}
\label{fig:characterizing}
\end{figure}

Next, we want to see what the characterizing movements for different instruments. That is, the movements that occur relatively more often in one instrument comparing to other instruments. For each of the nine instruments, we first collect the positive frames in the manually-annotated test data. Then, we compute the spatial maximum $[D_h]=\max_{a, b} [C_{h, a, b}]$ of the rectified output $[C_{h, a, b}]$ of Conv1 for each positive frame, where $h$ is the filter index and $(a, b)$ is the spatial coordinate. We compute the average over all $[D_h]$ of all positive frames to form a $96$-dimensional vector. There are nine instruments, so we get a $9\times 96$ matrix. Finally, we normalize the matrix along the first dimension and sort along the second dimension to get the top filters. The top five filters after this process are shown in Fig. \ref{fig:characterizing}.


We can see several interesting patterns in this visualization. The characterizing movements of Accordion are all horizontal. Cello has mostly smooth movements but also contains some twisting in some movements. Drum and Guitar have more back-and-forth patterns, indicating that they have faster movements than the others. Most of the movements in Violin are smooth, similar to those in Cello.


\subsection{Analyses and observations}
In this section, we share some analyses and observations we have made in the process of doing experiments. 

From TABLE \ref{tab:action_classwise}, we can see that the Drum action has moderate temporal performance but the worst spatial performance among the nine instruments. In other words, the model can usually correctly detect the timing of drum playing, but predict at the wrong spatial location. We find this result sensible because the contact of the stick and the drum skin usually lasts for a very short period of time. After the contact, the stick would go somewhere else rapidly. Therefore, it is too fast for the model to correctly locate the contact point.

It is relatively more difficult to detect the playing timing of instruments in the wind family, such as Flute, Saxophone, and Trumpet, because the playing actions of these instruments are usually subtle finger movements. This is also reflected by the temporal performance shown in TABLE \ref{tab:action_classwise}. Flute has another disadvantage due to its thin body. The object model often cannot well detect the Flute object, and this in turn affects the performance of the action models trained with the object model. In fact, Piano is another instrument that sometimes shares this difficulty because piano playing actions sometimes contain only finger movements while the hands do not move. We believe this is one of the reasons that the model does not perform well temporally on Piano.

Another interesting detection error of piano playing is from the optical reflection of hands on the polished cover of the pianos. We observe in several videos that the hands are reflected on the shiny black cover and the movements of the hands are also reflected on the cover. Both the object model and the action model can make false positives due to the optical reflection.

Originally, we thought that the playing actions of violin and cello might be similar and the model might have difficulty distinguishing them. Surprisingly, they are usually detected correctly without confusion.

We have seen ego-motions of cameras in many instrument-playing videos, and their effect to the action model would depend on several factors. First of all, we found that the music-related videos with ego-motions are usually filmed under worse conditions with poor resolutions because they are made by less professional people and in less formal occasions. Therefore, the problem of ego-motions is often coupled with worse video resolutions and serious occlusions. When these ill conditions come together, the model usually fails. 

Although we do not consider the ego-motion when we conduct the experiments, the model could handle the ego-motion in some circumstances. Assume we are in a scene where the player is not playing the instrument and there are ego-motions. We can handle the ego-motions in this situation if we use the fused prediction that combines the information of the sound activation into the action prediction, such as Action$\times$Sound or Action$\times$Object$\times$Sound discussed in Section \ref{sec:fusion_method}, because the action activations caused by ego-motions will be suppressed by the low activations of the sound model.

\section{Conclusions} \label{sec:conclusion}
In this paper, we have proposed a weakly-supervised framework to train a model for detecting instrument-playing actions in videos. In this framework, an auxiliary sound model and an auxiliary object model are utilized to provide supervisions to alleviate the lack of annotated data. We have shown that the proposed framework can significantly improve the detection ability both temporally and spatially.

There are several possible future directions. First, subtle finger movements are important cues for playing actions of many instruments, and the current model might not be able to fully capture them. A possible way to improve it is to automatically detect hand gestures or mouth movements and use them as input features. Second, the proposed framework could also be applied to other categories of actions where sounds, objects, or both are important cues of the actions.

\appendices

\section{Possible alternative models for the object model} \label{sec:fasterrcnn}

\begin{table*}[htbp]
	\renewcommand{\arraystretch}{1.3}
	\caption{The performance of the action models trained with a Faster R-CNN object model in comparison with the action models trained with a weakly-supervised object model. We train the action models with five thresholds in the object model: 0.1, 0.3, 0.5, 0.7, and 0.9. We let them use their own best threshold for each instrument because they are quite different models so they may have different optimal thresholds. We use 0.5 as the threshold in the sound model for SOT models.}
	\label{tab:wslvsrcnn}
	\centering
	\begin{tabular}{c|c|l||ccccccccc||c}
		\hline
		& \multirow{2}{*}{Action model}	& \multirow{2}{*}{Object model}	& Accordion	&Cello		&Drum		& Flute	& Guitar	& Piano	& Saxophone	& Trumpet	& Violin	& \multirow{2}{*}{Average}	\\
		&		&	& action	& action	& action	& action	& action	& action	& action	& action	& action	&	\\
		\hline
		\multirow{4}{*}{\shortstack{Temporal\\(AUC)}}	& \multirow{2}{*}{OT}	& Weakly-supervised				& \textbf{0.890}	&	\textbf{0.932}	&	0.831	&	\textbf{0.756}	&	0.788	&	0.722	&	\textbf{0.751}	&	\textbf{0.772}	&	\textbf{0.891}	&	\textbf{0.815} 		\\
		&						& Faster R-CNN 		& 0.835	&	0.913	&	\textbf{0.858}	&	0.689	&	\textbf{0.819}	&	\textbf{0.761}	&	0.737	&	0.751	&	0.883	&	0.805		\\
		\cline{2-13}
		& \multirow{2}{*}{SOT}	& Weakly-supervised				& \textbf{0.894}	&	\textbf{0.937}	&	0.854	&	\textbf{0.785}	&	\textbf{0.864}	&	0.778	&	\textbf{0.754}	&	0.777	&	\textbf{0.904}	&	\textbf{0.838}  		\\
		&						& Faster R-CNN 		& 0.859	&	0.921	&	\textbf{0.886}	&	0.744	&	0.860	&	\textbf{0.791}	&	0.753	&	\textbf{0.796}	&	0.890	&	0.833		\\
		\hline
		\multirow{4}{*}{\shortstack{Spatial\\(Pixel)}}	& \multirow{2}{*}{OT}	& Weakly-supervised				& 24.0	&	\textbf{23.1}	&	\textbf{40.3}	&	\textbf{26.4}	&	29.2	&	37.2	&	\textbf{23.2}	&	\textbf{34.3}	&	\textbf{17.8}	&	\textbf{28.4} 		\\
		&						& Faster R-CNN 		& \textbf{23.0}	&	23.5	&	45.3	&	31.4	&	\textbf{25.7}	&	\textbf{33.6}	&	23.9	&	38.1	&	21.0	&	29.5		\\
		\cline{2-13}
		& \multirow{2}{*}{SOT}	& Weakly-supervised				& \textbf{23.2}	&	\textbf{24.6}	&	\textbf{40.7}	&	\textbf{26.2}	&	27.3	&	\textbf{31.9}	&	\textbf{25.0}	&	35.5	&	\textbf{17.7}	&	\textbf{28.0} 		\\
		&						& Faster R-CNN 		& 26.1	&	26.7	&	42.8	&	31.6	&	\textbf{27.1}	&	42.1	&	26.1	&	\textbf{34.2}	&	21.3	&	30.9		\\
		\hline
		
	\end{tabular}
	
\end{table*}

In this section, we discuss possible alternative models for the object model. We will refer to the object model we have used so far as the weakly-supervised object model in this section.

In this paper, we want the object model to locate instruments in a frame or an image. There are mainly two types of detection in the literature: instance segmentation for pixel-level prediction \cite{he2017} and object detection for bounding-box prediction \cite{girshick2015, ren2015}. In our application, we prefer the pixel-level prediction because a state-of-the-art model (such as Mask R-CNN \cite{he2017}) can segment the objects precisely. In contrast, bounding-box could include undesired regions external to the target object. However, a model for pixel-level prediction usually requires training with pixel-level annotations \cite{he2017} which are expensive to collect. Microsoft COCO \cite{lin2014} is a commonly used dataset for instance segmentation \cite{he2017}. It contains 91 classes of objects, but contains no instrument classes. Therefore, we are not able to use it to train instance segmentation models.

To address this problem, we may predict the pixel-level labels by using weakly supervised learning as it is done by Oquab \emph{et al.} \cite{oquab2015} and as it is done in this paper. We may also predict the bounding boxes by using bounding-box annotations, which are easier to collect than the pixel-level annotations. PASCAL VOC datasets are commonly used datasets for bounding-box-based object detection\footnote{http://host.robots.ox.ac.uk/pascal/VOC/index.html} \cite{everingham10}, which contain images of 20 classes with bounding-box information, but the classes do not contain instruments either. Nevertheless, we can train an object detection model by using ImageNet-Instrument data we have collected from the ImageNet website, which have been used to evaluate the object model as described in Section \ref{sec:imagenet_instrument}.

In the rest of this section, we investigate using Faster R-CNN as an object model for training OT and SOT models. We conduct this experiment by modifying the code from https://github.com/ruotianluo/pytorch-faster-rcnn and re-organizing the data of ImageNet-Instrument.

\begin{figure}[!t]
	\centering
	\includegraphics[width=0.95\columnwidth]{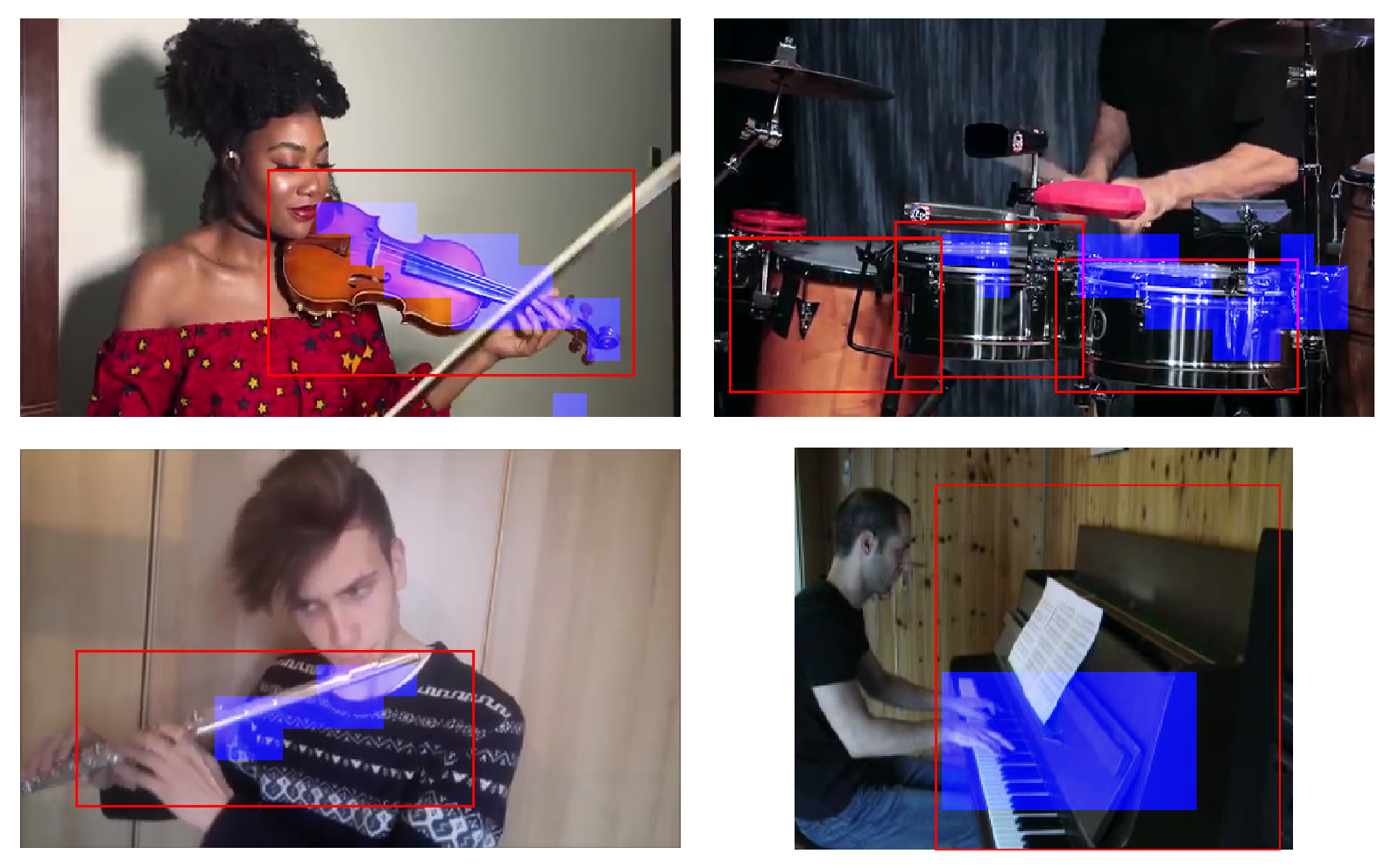}
	\caption{Prediction of the weakly-supervised object model used in this paper and the prediction of a Faster R-CNN model. The red rectangle frames are the predictions of the Faster R-CNN model, and the blue shades are the predictions of the weakly-supervised object model. Best seen in color.}
	\label{fig:objectmodels}
\end{figure}

The Faster R-CNN object model achieves $96.8 \%$ hit rate if we evaluate it in the way we evaluate the object model in Section \ref{sec:object_eval}. We already use the ImageNet-Instrument data for training the Faster R-CNN model, so this is not a fair comparison. Nonetheless, we also test the Faster R-CNN model on several external images containing instruments and find it performs very well. Some examples of the predicted bounding boxes of the Faster R-CNN model are shown in Fig. \ref{fig:objectmodels}. The red bounding boxes are the predictions of the Faster R-CNN model, and we show the predictions of the weakly-supervised object model alongside with blue shades. We can see that the Faster R-CNN predictions are pretty accurate, but the bounding boxes also contain large amount of parts that are not instruments or are not relevant to making instrument sounds due to the shape of the bounding boxes. In contrast, the predictions of the weakly-supervised object model often fail to cover the entire instrument, but they fit better the shape of the instruments. 

We use the trained Faster R-CNN as the object model and train OT and SOT models by applying five thresholds (0.1, 0.3, 0.5, 0.7, and 0.9) to the class scores of Faster R-CNN. As the weakly-supervised object model and the Faster R-CNN object model are very different models, they may have different optimal thresholds. Therefore, we use their own best threshold for each instrument in order to compare their performance. The result is shown in TABLE \ref{tab:wslvsrcnn}. 

Despite of the good performance of Faster R-CNN in predicting bounding boxes, we find that training SOT and OT action models with it does not yield better performance. 
In terms of the temporal performance of playing-action detection, the action models trained with the weakly-supervised object models and Faster R-CNN are close. We can see that the Faster R-CNN object model is better at Drum, Guitar and Piano for the OT model and is better at Drum and Piano for the SOT model. Either OT or SOT gets better result for Drum and Piano with the Faster R-CNN object model. It is interesting that Drum and Piano happen to be the two non-portable instruments among the nine instruments due to their sizes (cf. TABLE \ref{tab:instrument}). 

Our conjecture is that this difference in Drum and Piano actions results from the ratio of the playing region of an instrument to the size of the whole instrument. For Drum and Piano, the playing regions are relatively small compared to the size of the entire instruments. As we can see from Fig. \ref{fig:objectmodels}, the weakly-supervised object model only predicts as positive a small portion of the piano and drums, which will make it more difficult to intersect with hands or sticks. In contrast, Faster R-CNN can very nicely recognize the entire instrument which will also include relevant body motions other than the playing actions. Therefore, Faster R-CNN could better predict the playing actions temporally in Drum and Piano in comparison to the weakly-supervised object model.

This argument is also supported by the spatial performance of the action models trained with Faster R-CNN. The larger region predicted by Faster R-CNN model will also produce more false positives spatially. Therefore, the average spatial performance of the action models trained with Faster R-CNN is worse than that with the weakly-supervised object model, and it even gets worse in Drum and Piano.



%

\bibliographystyle{IEEEtran}
\bibliography{jy17}

%


\begin{IEEEbiographynophoto}
{Jen-Yu Liu}
is a Ph.D. student in the Electrical Engineering Department in National Taiwan University and a research assistant in the Research Center for Information Technology Innovation in Academia Sinica, Taiwan. His research interests include music auto-tagging, instrument detection, weakly-supervised learning, and topological data analysis.
\end{IEEEbiographynophoto}
\begin{IEEEbiographynophoto}
{Yi-Hsuan Yang} (M’11--SM'17)
is an Associate Research Fellow with Academia Sinica. He received his Ph.D. degree in Communication Engineering from National Taiwan University in 2010.
He is also a Joint-Appointment Associate Professor with the National Cheng Kung University, Taiwan. His research interests include music information retrieval, affective computing, multimedia, and machine learning. Dr. Yang was a recipient of the 2011 IEEE Signal Processing Society Young Author Best Paper Award, and the 2015 Best Conference Paper Award of the IEEE Multimedia Communications Technical Committee. He is an author of the book Music Emotion Recognition (CRC Press 2011).  In 2014, he served as a Technical Program Co-Chair of the International Society for Music Information Retrieval Conference (ISMIR). In 2016, he started his term as an Associate Editor for the \emph{IEEE Transactions on Affective Computing} and the \emph{IEEE Transactions on Multimedia}. Dr. Yang is a senior member of the IEEE.
\end{IEEEbiographynophoto}
\begin{IEEEbiographynophoto}
{Shyh-Kang~Jeng}
In 1981 he joined the faculty of the Department of Electrical Engineering, National Taiwan University, where he is now a Professor. From 1984 to 1985 he was an electronic data processing officer and an instructor on information system analysis and design at the National Defense Management College, Chung-Ho, Taiwan, R.O.C. From 1985 to 1993 he visited University of Illinois, Urbana-Champaign, USA, as a Visiting Research Associate Professor and a Visiting Research Professor several times. In 1999 he visited Center for Computer Research in Music and Acoustics, Stanford University, USA, for half of a year. He also served as a Session Chairman in 1994 Joint International IEEE/APS Symposium and URSI Radio Science Meeting in Seattle, USA, and 2005 IEEE AP-S International Symposium and USNC/URSI Radio Science Meeting in Washington DC, USA. He has also been invited to review papers for IEEE Transactions on Antennas and Propagation, IEEE Transactions on Microwave Theory and Techniques, IEEE Transactions on Vehicular Technology, and IEEE Transactions on Multimedia. He is also a recipient of the 1998 Outstanding Research Award of National Science Council and 2004 Outstanding Teaching Award of National Taiwan University. His research interest includes theory and applications of electromagnetics, music signal processing, computational cognitive neuroscience, and cognitive neurorobotics. 
\end{IEEEbiographynophoto}






\end{document}